\documentclass[twocolumn, twocolappendix]{aastex631}
\usepackage[T5]{fontenc}
\usepackage[utf8]{inputenc}
\usepackage{svg}
\usepackage{gensymb}
\usepackage{amsmath}
\usepackage{rotating}
\def\msun{\,{\rm M_\odot}}

\usepackage[dvipsnames]{xcolor}
\definecolor{mycyan}{rgb}{0.5, 1.0, 1.0}
\definecolor{myorange}{rgb}{1.0, 0.824, 0.5}
\definecolor{myred}{rgb}{1.0, 0.5, 0.5}
\definecolor{mydarkred}{rgb}{0.773, 0.5, 0.5}

\defcitealias{Kim+2013}{I}
\defcitealias{Kim+2016}{II}
\defcitealias{Roca-Fabrega+2021}{III}
\defcitealias{Roca-Fabrega+2024}{IV}
\defcitealias{Jung+2024}{V}
\defcitealias{Strawn+2024}{VI}
\defcitealias{Rodriguez-Cardoso+2025a}{VII}
\defcitealias{Jung+2025a}{VIII}
\defcitealias{Kim+2025}{X}
\defcitealias{Barrow+2026a}{XI}

\begin{document}

\title{The \textit{AGORA} High-resolution Galaxy Simulations Comparison Project. IX - Part 2: Effects of a Major Galaxy Merger on the Stellar Morphology of a Milky Way-mass Galaxy Progenitor}
\author[0009-0002-2290-8039]{Thịnh Hữu Nguyễn}
\altaffiliation{Corresponding author}
\affiliation{Department of Astronomy, University of Illinois at Urbana-
Champaign, Urbana, IL 61801, USA; thinhhn2@illinois.edu}
\affiliation{Center for AstroPhysical Surveys, National Center for Supercomputing Applications, Urbana, IL 61801, USA}

\author[0000-0002-8638-1697]{Kirk S. S. Barrow}
\affiliation{Department of Astronomy, University of Illinois at Urbana-
Champaign, Urbana, IL 61801, USA}

\author[0000-0002-9144-1383]{Minyong Jung}
\altaffiliation{Corresponding author}
\affiliation{Center for Theoretical Physics, Department of Physics and Astronomy, Seoul National University, Seoul 08826, Republic of Korea; wispedia@snu.ac.kr}

\author[0000-0002-9158-195X]{Ram\'{o}n Rodr\'{i}guez-Cardoso}
\altaffiliation{Corresponding author}
\affil{Departamento de Física de la Tierra y Astrofísica, Fac. de C.C. Físicas, Universidad Complutense de Madrid, E-28040 Madrid, Spain; ramorodr@ucm.es}
\affil{Instituto de Física de Partículas y del Cosmos, IPARCOS, Fac. C.C. Físicas, Universidad Complutense de Madrid, E-28040 Madrid, Spain}

\author[0000-0002-6299-152X]{Santi Roca-F\`{a}brega}
\altaffiliation{Code leaders}
\affil{Lund Observatory, Division of Astrophysics, Department of Physics, Lund University, SE-221 00 Lund, Sweden}
\affil{Departamento de F\'{i}sica de la Tierra y Astrof\'{i}sica, Facultad de Ciencias F\'{i}sicas, Plaza Ciencias, 1, 28040 Madrid, Spain}

\author[0000-0003-4464-1160]{Ji-hoon Kim}
\altaffiliation{Code leaders}
\affiliation{Seoul National University Astronomy Research Center, Seoul 08826, Korea}
\affiliation{Center for Theoretical Physics, Department of Physics and Astronomy, Seoul National University, Seoul 08826, Korea}
\affiliation{Institute for Data Innovation in Science, Seoul National University, Seoul 08826, Republic of Korea}

\author[0000-0001-5091-5098]{Joel R. Primack}
\altaffiliation{Deceased}
\affil{Department of Physics, University of California at Santa Cruz, Santa Cruz, CA 95064, USA}

\author[0000-0001-7457-8487]{Kentaro Nagamine}
\altaffiliation{Code leaders}
\affiliation{Theoretical Astrophysics, Department of Earth and Space Science, Graduate School of Science, Osaka University, Toyonaka, Osaka, 560-0043, Japan}
\affiliation{Theoretical Joint Research, Forefront Research Center, Graduate School of Science, The University of Osaka, 1-1 Machikaneyama, Toyonaka, Osaka 560-0043, Japan}
\affiliation{Kavli IPMU (WPI), UTIAS, The University of Tokyo, Kashiwa, Chiba 277-8583, Japan}
\affiliation{Department of Physics \& Astronomy, University of Nevada Las Vegas, Las Vegas, NV 89154, USA}
\affiliation{Nevada Center for Astrophysics, University of Nevada, Las Vegas, 4505 S. Maryland Pkwy, Las Vegas, NV 89154-4002, USA}

\author[0000-0001-8531-9536]{Renyue Cen}
\affil{Center for Cosmology and Computational Astrophysics, Institute for Advanced Study in Physics, Zhejiang University, Hangzhou 310027, People's Republic of China}
\affil{Institute of Astronomy, School of Physics, Zhejiang University, Hangzhou 310027, People's Republic of China}

\author[0000-0002-8680-248X]{Daniel Ceverino}
\affil{Universidad Aut\'{o}noma de Madrid, Ciudad Universitaria de Cantoblanco, E-28049 Madrid, Spain}
\affil{CIAFF, Facultad de Ciencias, Universidad Aut\'{o}noma de Madrid, E-28049 Madrid, Spain}

\author[0000-0002-2113-4863]{Weiguang Cui}
\affil{Institute for Astronomy, Royal Observatory, Edinburgh EH9 3HJ, UK}
\affil{Departamento de Física Teórica, Universidad Autónoma de Madrid, Módulo 15, E-28049 Madrid, Spain}
\affil{Centro de Investigación Avanzada en Física Fundamental (CIAFF), Facultad de Ciencias, Universidad Autónoma de Madrid, E-28049 Madrid, Spain}

\author[0000-0003-0073-3012]{Anna Genina}
\altaffiliation{Code leaders}
\affil{Institute for Astronomy, University of Edinburgh, Royal Observatory, Blackford Hill, Edinburgh EH9 3HJ, UK}

\author[0000-0002-7820-2281]{Hyeonyong Kim}
\altaffiliation{Code leaders}
\affiliation{Center for Theoretical Physics, Department of Physics and Astronomy, Seoul National University, Seoul 08826, Korea}

\author[0000-0002-5712-6865]{Yuri Oku}
\affiliation{Theoretical Astrophysics, Department of Earth and Space Science, Graduate School of Science, Osaka University, Toyonaka, Osaka, 560-0043, Japan}
\affil{Center for Cosmology and Computational Astrophysics, Institute for Advanced Study in Physics, Zhejiang University, Hangzhou 310027, People's Republic of China}

\author[0000-0002-3764-2395]{Johnny W. Powell}
\altaffiliation{Code leaders}
\affil{Department of Physics, Reed College, Portland, OR 97202, USA}

\author[0000-0002-6227-0108]{Yves Revaz}
\altaffiliation{Code leaders}
\affil{Institute of Physics, Laboratoire d'Astrophysique, \'{E}cole Polytechnique F\'{e}d\'{e}rale de Lausanne (EPFL), CH-1015 Lausanne, Switzerland}

\author[0009-0002-1398-6537]{Pablo Granizo}
\affiliation{Theoretical Astrophysics, Department of Earth and Space Science, Graduate School of Science, Osaka University, Toyonaka, Osaka, 560-0043, Japan}
\affil{Universidad Aut\'{o}noma de Madrid, Ciudad Universitaria de Cantoblanco, E-28049 Madrid, Spain}

\author[0000-0001-6106-7821]{Alessandro Lupi}
\altaffiliation{Code leaders}
\affil{Como Lake Center for Astrophysics, DiSAT, Universit\`a degli Studi dell'Insubria, via Valleggio 11, I-22100 Como, Italy}
\affil{INFN, Sezione di Milano-Bicocca, Piazza della Scienza 3, I-20126 Milano, Italy}
\affil{INAF - Osservatorio di Astrofisica e Scienza dello Spazio di Bologna, Via Gobetti 93/3, I-40129 Bologna}

\author{Ikkoh Shimizu}
\altaffiliation{Code leaders}
\affil{Shikoku Gakuin University, 3-2-1 Bunkyocho, Zentsuji, Kagawa, 765-8505, Japan}

\author{H\'{e}ctor Vel\'{a}zquez}
\altaffiliation{Code leaders}
\affil{Instituto de Astronom\'{i}a, Universidad Nacional Aut\'{o}noma de M\'{e}xico, A.P. 70-264, 04510, Mexico, D.F., Mexico}

\author[0000-0002-5969-1251]{Tom Abel}
\affil{Kavli Institute for Particle Astrophysics and Cosmology, Stanford University, Stanford, CA 94305, USA}
\affil{Department of Physics, Stanford University, Stanford, CA 94305, USA}
\affil{SLAC National Accelerator Laboratory, Menlo Park, CA 94025, USA}

\author[0000-0002-4287-1088]{Oscar Agertz}
\affil{Lund Observatory, Division of Astrophysics, Department of Physics, Lund University, SE-221 00 Lund, Sweden}

\author[0000-0003-4174-0374]{Avishai Dekel}
\altaffiliation{Deceased}
\affil{Center for Astrophysics and Planetary Science, Racah Institute of Physics, The Hebrew University, Jerusalem 91904, Israel}

\author[0000-0003-4597-6739]{Boon Kiat Oh}
\affiliation{Department of Physics, University of Connecticut, U-3046, Storrs, CT 06269, USA}
\affiliation{School of Physics, Korea Institute for Advanced Study, 85 Hoegiro, Dongdaemun-gu, Seoul 02455, Republic of Korea}

\author[0000-0001-5510-2803]{Thomas R. Quinn}
\affil{Department of Astronomy, University of Washington, Seattle, WA 98195, USA}

\author{the {\it AGORA} Collaboration}



\begin{abstract}

Galaxy mergers, with their high sensitivity to initial conditions, provide a valuable setting for comparative studies of galaxy simulation codes. Following our first paper focusing on merger-driven star formation, we present a code comparison examining the morphological transformation impact of a major galaxy merger at $z \approx 4.5$ on a Milky Way-mass galaxy progenitor. Our analysis employs nine state-of-the-art codes from the AGORA \texttt{CosmoRun} cosmological zoom-in simulation suite. For this merger, we show that the adopted stellar feedback type influences the galaxy's compaction and stellar disc formation. Codes with purely thermal feedback produce a merger remnant that forms a disc and becomes compact primarily during and after coalescence; codes that include kinetic feedback begin disc formation and compaction around the first periapsis; and codes with strong delayed cooling or superbubble feedback suppress disc formation and produce a more extended remnant. In contrast, the orientation of the remnant disc is code-independent. In all codes, the rotational angular momentum of the remnant disc aligns with the interaction's orbital angular momentum rather than the pre-merger rotational axis, implying that the infalling gas preserves its orbital angular momentum to form a new disc. Comparisons with the Santa Cruz semi-analytic model show reasonable agreement in stellar mass and half-mass radius, yet the model underpredicts (overpredicts) the dark matter fraction and velocity dispersion for codes exhibiting strong compaction (expansion). The systematic dependence of our remnants' morphology on feedback schemes demonstrates that merger remnant morphology may serve as a powerful probe of stellar feedback processes. 

\end{abstract}




\section{Introduction} \label{sec:intro}

In the local Universe, the visual morphologies of luminous galaxies are broadly described by the Hubble sequence \citep{Hubble+1926}, which broadly classifies galaxies into early-type (elliptical and lenticular), late-type (spiral), and irregular categories. The relative fraction of each morphological category in this classification scheme, however, is not constant and evolves with redshift. By visually classifying galaxies from the SDSS, POWIR/DEEP2, and GOODS NICMOS surveys, \cite{Buitrago+2013} showed that, at stellar mass $M_\bigstar > 10^{11} M_\odot$, the fraction of early-type galaxies rises from~$20\%$ at $z \approx 3$ to~$70\%$ at $z \approx 0$, while the fraction of irregular galaxies declines over the same interval. \cite{Mortlock+2013} analyzed galaxies with $M_\bigstar > 10^{10} M_\odot$ at $z=1\text{--}3$ in the Hubble Space Telescope CANDELS survey, and found irregular morphology to be dominant at $z \geq 2$. Despite probing a broader mass range than \cite{Buitrago+2013}, they likewise reported that the spheroidal fraction increases with decreasing redshift. More recently, with the more advanced imaging capability from JWST, \cite{Kajisawa+2026} revisited this question with galaxies at $z = 0.2\text{--}2$ and $M_\bigstar > 10^{11} M_\odot$ using the COSMOS-Web survey. They also arrived at a similar trend for the fraction of early-type galaxies. This evolution of galaxy morphology with redshift implies that there exists a morphological transformation as galaxies evolve, and elucidating the responsible mechanisms is thus fundamental to understanding galaxy evolution. 

Galaxy mergers have been proposed as an important mechanism for transforming galaxy morphology. Several theoretical studies have explored this question using different methods and simulations. \cite{Hopkins+2010a} employed a semi-empirical model by combining observationally constrained halo occupation distributions with the halo–halo merger rate to predict the galaxy-galaxy merger rate. Then, the galaxy-galaxy merger rate was coupled with fitting functions calibrated with hydrodynamics simulations to compute the relative contributions of mergers to bulge formation in galaxies. Their results indicated that about 70\% of the globally integrated bulge mass originates from major galaxy mergers, while the other 30\% comes from minor mergers. They also found that early-type and lenticular galaxies preferentially emerge from major mergers, whereas later-type galaxies are more commonly formed from minor mergers. \cite{Wilman+2013} further analysed different independently developed semi-analytic models and also suggested that observed elliptical galaxies form during major mergers. Yet, they found that these models overestimate the elliptical fraction compared to the observed local universe. 

On the hydrodynamics simulation side,  using the Horizon-AGN cosmological simulation, \cite{Martin+2018} arrived at a similar conclusion, demonstrating that mergers with a mass ratio greater than 1:10 contribute substantially to the formation of massive spheroids ($M_\bigstar > 10^{10.5} M_\odot$) at z = 0. Moreover, minor mergers also play an important role in the transformation to massive spheroids, contributing one-third of the global transformation and becoming the dominant driver at $z \leq 1$. While similarly confirming that the morphology of galaxies with $M_\bigstar > 10^{11} M_\odot$ is largely influenced by mergers, the Illustris simulation shows that the morphology of small to medium-sized galaxies ($10^{9} M_\odot < M_\bigstar < 10^{10} M_\odot$) has less dependence on the merger history \citep{Rodriguez-Gomez+2017}. Another approach to studying mergers is the genetic modification technique in cosmological simulations, which enables controlled studies on the causal effect of a galaxy's merger history while preserving the galaxy's final halo mass and realistic large-scale environment \citep{Roth+2016}. By varying the stellar mass ratio (from $\approx$ 1:25 to 1:2) of an important merger at $z = 2$ of a Milky-Way-mass galaxy, \cite{Rey+2023} and \cite{Joshi+2025} showed that a smaller merger mass ratio produces a more extended, rotation-dominated galaxy at $z = 0$. In contrast, a higher mass ratio yields a more compact, bulge-dominated remnant. Furthermore, the outcome of any given merger is not only determined by its mass ratio and mass range, but also depends on the gas content of the progenitors. Early work by \cite{Springel+2005c} demonstrated that sufficiently gas-rich major mergers can reform a rotationally supported disc in the remnant rather than producing an elliptical galaxy, as gas that is not consumed in the merger-driven starburst cools and settles into an extended star-forming disc. \cite{Hopkins+2009} formalised this behaviour in a semi-analytic model that can predict the fraction of gas that survives to re-form a disc, the fraction of stars that undergo violent relaxation, and the fraction of gas that dissipatively loses angular momentum to fuel a central starburst. They found that the gas fraction plays a crucial role in determining the surviving disc fraction after a merger. This conclusion has been validated using galaxy mergers in cosmological simulations, including Illustris \citep{Peschken+2020, Rodriguez-Gomez+2017}, IllustrisTNG \citep{Pallero+2025}, and Horizon-AGN \citep{Martin+2018}. In addition to the gas fraction, the model of \citet{Hopkins+2009} is also a function of the progenitor galaxies' disc fraction, the orbital parameters, and the scale length of the disc stars. Moreover, the presence of a hot gaseous halo in a binary galaxy merger simulation can decrease the bulge-to-total ratio of the merger remnant, as the halo serves as a reservoir to replenish the disc with infalling cold gas \citep{Kannan+2015}. Orbital configuration adds an additional layer of complexity. For instance, polar and retrograde-prograde major mergers result in the highest asymmetry of the galaxy system during the interaction, while retrograde-retrograde major mergers have the longest timescale of disturbed morphology \citep{Lotz+2008}. \cite{Martin+2018} found that retrograde mergers can induce around twice the degree of morphological changes compared to the prograde counterparts at $z \approx 0$. 

Taken together, these results establish that merger-induced morphological transformation is governed by multiple physical parameters. Nevertheless, the degree to which a simulation code architecture and/or a feedback prescription influence the morphological impact of galaxy mergers remains relatively unexplored. Given that the work done on this topic has utilised multiple different simulation codes, this question becomes even more relevant. An answer to this question requires a carefully calibrated suite of cosmological simulations with a resolved galaxy merger. 

The AGORA (Assembling Galaxies of Resolved Anatomy) project \citep{Kim+2013, Kim+2016} is a forefront code comparison project that robustly calibrates nine state-of-the-art galaxy simulation codes so that the differences in the results can be directly tied to only a few variables. The simulations share common initial conditions and multiple key astrophysics packages, such as gas cooling/heating, star formation prescriptions, and UV background, while differing only in the numerical architecture and the stellar feedback subgrid model. \texttt{CosmoRun} is a fully cosmological zoom-in galaxy formation simulation suite developed by the AGORA collaboration~\citep[hereafter Papers \citetalias{Roca-Fabrega+2021} and \citetalias{Roca-Fabrega+2024}]{Roca-Fabrega+2021, Roca-Fabrega+2024}. Since it was created, the \texttt{CosmoRun} suite has been enabling comparative studies across various astrophysical contexts, including satellite galaxies~\citep[hereafter Paper \citetalias{Jung+2024}]{Jung+2024}, the circumgalactic medium~\citep[hereafter Paper \citetalias{Strawn+2024}]{Strawn+2024}, the mechanisms of satellite quenching~\citep[hereafter Paper \citetalias{Rodriguez-Cardoso+2025a}]{Rodriguez-Cardoso+2025a}, the formation and evolution of galactic disc~\citep[hereafter Paper \citetalias{Jung+2025a}]{Jung+2025a}, the dark matter halo morphology~\citep[hereafter Paper \citetalias{Barrow+2026a}]{Barrow+2026a}. In this work, we conduct a comparative study of a major galaxy merger happening to the main galaxy at $z=4.5$ in \texttt{CosmoRun}, and the study is divided into two parts. In the companion Paper~IX - Part~1 (Nguyen et al. submitted), we investigate the effect of the merger on star formation. This paper is the second part of the series, where we concentrate on the merger-driven morphological transformation, especially with regard to the stellar radial distribution and the disc formation.  Similar to Paper~IX - Part~1, we refer to the major galaxy merger at $z=4.5$ that we analyse as \textbf{the \textit{target} merger} to avoid confusion with other galaxy merger events in the simulations.

The paper is organised as follows. In Section~\ref{sec:method}, we describe the technical aspects of the AGORA \texttt{CosmoRun} simulation suite, the merger tree, and our method to define merger timings, merger stages, and the galactic centre. In Section~\ref{sec:effect_on_morphology}, we discuss the changes in the stellar radial distribution of the main galaxy after the merger (Section~\ref{subsect:radial_distribution}), the relationship between burst fraction and the change in the galaxy's half-mass radius (Section~\ref{subsect:burstfraction_vs_compaction}), and disc formation after the merger (Section~\ref{subsect:disk_formation}). We compare our results with the Santa-Cruz semi-analytic model and observations in Section~\ref{sec:discussion}. Finally, Section~\ref{sec:conclusion} summarises the paper and discusses some implications of the findings. 

\section{Methods} \label{sec:method}

\subsection{The AGORA \texttt{CosmoRun} Simulation Suite}

The AGORA \texttt{CosmoRun} is a suite of high-resolution cosmological zoom-in hydrodynamic simulations of a Milky-Way–mass halo ($\text{M}_\text{halo} \approx 10^{12}\;\text{M}_\odot$ at $z = 0$) across multiple simulation codes. The \texttt{CosmoRun} suite currently has nine codes: three adaptive mesh refinement (AMR) codes, ART-I~\citep{Kravtsov+1997}, ENZO~\citep{Bryan+2014}, and RAMSES~\citep{Teyssier+2002}; four particle-based codes, CHANGA~\citep{Jetley+2008a, Jetley+2010, Menon+2015}, GADGET-3 (GADGET3-OSAKA version; \citealp{Shimizu+2019, Nagamine+2021}; base code by \citealp{Springel+2005}), GADGET-4 (GADGET4-OSAKA version; \citealp{Romano+2022, Romano+2022a, Oku+2024, Granizo+2026}; base code by \citealp{Springel+2021}), and GEAR (\citealp{Revaz+2012}; base code by \citealp{Springel+2005}); one moving-mesh code, AREPO~\citep{Springel+2010, Weinberger+2020} (the version of AREPO used in this paper is referred to as AREPO‑T, indicating the use of only thermal feedback for its stellar feedback prescription); and one meshless/hybrid code, GIZMO~\citep{Hopkins+2015}. All simulations started from $z = 100$ and reached at least $z = 2$, with several codes continuing to $z \approx 0$. The choice to evolve to lower redshifts was left to each code group and does not reflect the computational cost or the code performance. Each simulation is calibrated so that its main halo achieves a stellar mass at $z=4$ that agrees with predictions from semi-empirical models ($M_\bigstar = 1\text{--}5\times10^{9}\msun$). We assume a flat $\Lambda$CDM cosmology with WMAP7/9+SNe+BAO cosmological parameters~\citep{Komatsu+2011, Hinshaw+2013}: $\Omega_{m} = 0.272$, $\Omega_{\Lambda} = 0.728$, $h = 0.702$, $\sigma_{8} = 0.807$, and $n_{s} = 0.961$. 

All \texttt{CosmoRun} simulations begin from the same initial conditions generated with \textsc{music} \citep{Hahn+2011}. The simulations have a comoving volume of~$60\,(\mathrm{Mpc}/h)^3$ with a~$128^3$ root grid. The zoom-in region is created using five additional levels of refinement, yielding an effective resolution of~$4096^{3}$ in the highest-resolution region. For the AMR codes, the finest AMR cell size is~163~comoving pc (12 additional refinement levels from the root resolution). Cells are refined when their baryon mass or the particle mass exceeds four times the subgrid's mean density. For particle-based and hybrid codes, the gravitational softening length in the highest-resolution region is~800~comoving pc for $z > 9$ and~80~proper pc thereafter. The highest mass resolution for dark matter (DM) and gas particles (for codes that use them) is $m_{\text{DM}, \text{IC}}=2.8\times10^{5}\msun$ and $m_{\text{gas}, \text{IC}}=5.65\times10^{4}\msun$, respectively. The simulations also share many common astrophysical packages, including the radiative gas cooling library \textsc{grackle} \citep{Smith+2017}, the redshift-dependent cosmic UV background \citep{Haardt+2012}, and the star formation criteria. Star formation occurs at a gas density threshold of $n_\text{H,thres}=1\, \text{cm}^{-3}$ ($n_\text{H}$ is the total hydrogen number density) and follows a rate of $d\rho_{\star}/dt = \epsilon_\star\rho_\text{gas}/t_{\text{ff}}$, where $\epsilon = 0.01$ is the formation efficiency and $t_\text{ff}$ is the local freefall time. In particle-based codes and GIZMO, the initial mass of a star particle is the mass of its parent gas particle. CHANGA, as an exception, uses a constant initial stellar mass of $5.65\times10^{4}\msun$. For more details about the initial conditions and common astrophysics package, we direct the readers to Section~2 of Paper~\citetalias{Kim+2013} and Section~2 of Paper~\citetalias{Roca-Fabrega+2021}. 

\begin{table}
\vspace*{1mm}
\caption{\footnotesize The general type of the stellar feedback implementation adopted by each code group.\tablenotemark{\textdagger}}
\centering
\begin{tabular}{c  c  c  c }
\hline\hline 
Code & Stellar feedback\\ 
\hline
{\sc Art-I} & T+K, RP  \\
{\sc Enzo} & T   \\ 
{\sc Ramses} & T, DC   \\
{\sc Changa} & T+S  \\  
{\sc Gadget-3} & T+K, RP, DC  \\
{\sc Gadget-4} & T+K  \\ 
{\sc Gear} & T, DC   \\
{\sc Arepo-T} & T   \\
{\sc Gizmo} & T+K  \\  
\hline 
\end{tabular}
\tablenotetext{$\textdagger$}{\scriptsize T = thermal feedback, K = kinetic feedback, RP = radiation pressure, DC = delayed cooling, S = superbubble. We note that the literature sometimes draws a distinction between kinetic and mechanical feedback \citep{Rosdahl+2017}. In the kinetic scheme, the injected momentum is governed by a free parameter, whereas in the mechanical scheme, it depends on whether the Sedov--Taylor phase of the supernova is resolved. ART-I and GADGET-3 employ the kinetic feedback, while GADGET-4 and GIZMO employ the mechanical feedback. For simplicity, since both schemes operate by injecting momentum, we refer to them collectively as "kinetic feedback" (K). The feedback scheme of each code group follows the prevalent practice in its community as closely as possible. For more details about the specific implementation, the readers can refer to Papers \citetalias{Roca-Fabrega+2021}, \citetalias{Roca-Fabrega+2024}, and Paper~IX - Part~1 and the references within.}
\label{tab:feedback}
\vspace*{2mm}
\end{table}

One of the key differences between the \texttt{CosmoRun} codes is the stellar feedback scheme. The stellar feedback prescription of each code follows closely the widely adopted one in its respective code community. Because \texttt{CosmoRun} was not run with AGN feedback, references to "feedback" hereafter denote stellar feedback. Table~\ref{tab:feedback} displays the general type of stellar feedback scheme each code employs. Readers may refer to Papers IX - Part 1, ~\citetalias{Roca-Fabrega+2021}, ~\citetalias{Roca-Fabrega+2024}, and \citetalias{Jung+2025a} for more details on the feedback model parameters. It is important to note that because other variations of stellar feedback prescriptions still exist in each code community, results from simulations that use AGORA codes but adopt non-\texttt{CosmoRun} feedback models should be compared to \texttt{CosmoRun} results with appropriate caution.

\subsection{Halo Finding and Stellar Assignment}
\label{subsect:halofinding}

The DM halos and the merger tree are identified using \textsc{HASKAP PIE} (\citealp{Barrow+2026}, Paper~\citetalias{Barrow+2026a}), an all-in-one algorithm that performs both halo finding and merger tree construction via overdensity-finding, energy-solving, cluster-finding, and particle tracking. \textsc{HASKAP PIE} defines halos using the DM particles' orbital energy, allowing robust definitions of non-spherical halos and ensuring that the halo structure comprises only mutually bound particles. The non-spherical shape of a \textsc{HASKAP PIE} halo is defined by a convex hull. The algorithm also calculates the orbital energy of each star particle with respect to all identified DM halos and assigns stars to the halos with which they have negative orbital energy. The details of the merger history of the main galaxy can be found in Paper~IX - Part~1.

\subsection{Merger timings and parameters}
\label{subsect:merger_timing_and_stages}

\begin{figure*}[tbh]
    \centering
    \includegraphics[width=0.949\linewidth]{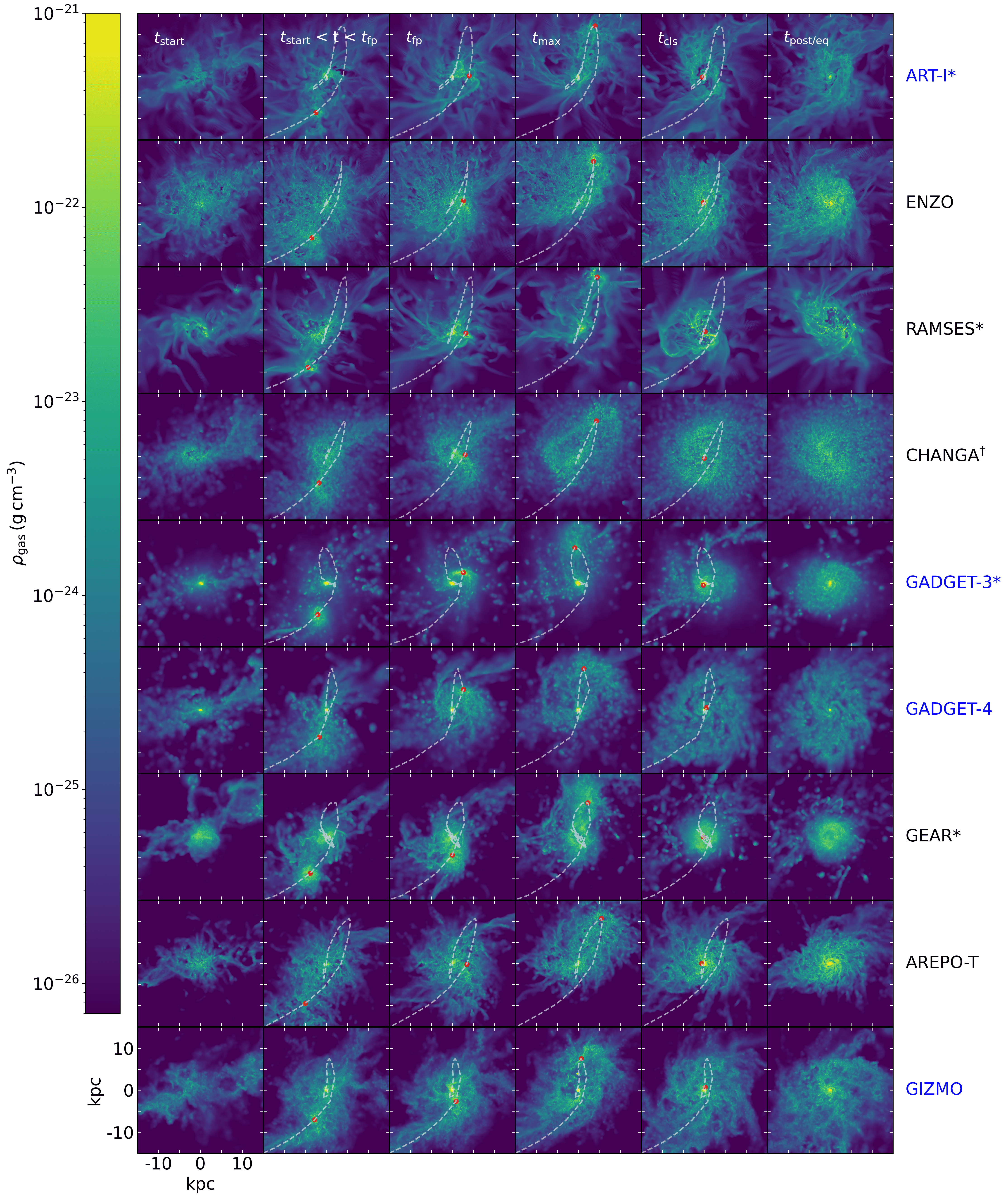}
    \caption{Gas projection plots showing key moments - $t_\text{start}$ (column 1), an intermediate timestep between  $t_\text{start}$ and $t_\text{fp}$ (column 2), $t_\text{fp}$ (column 3), $t_\text{max}$ (column 4), $t_\text{cls}$ (column 5), and $t_\text{post/eq}$ (column 6) - of the \textit{target} merger across nine \texttt{CosmoRun} codes. For each code, all projections are centred on the primary galaxy's centre with their normal vector aligned with the system's angular momentum at $t_\text{post/eq}$ (face-on view at $t_\text{post/eq}$). For visual references, the trajectory and the location of the secondary galaxy at each timestep are overplotted as dashed white lines and red dots, respectively. Blue code names indicate codes with kinetic feedback, the asterisk sign (*) denotes codes with delayed cooling and/or radiation pressure schemes, and the dagger sign (†) denotes codes with the superbubble scheme.}
    \label{fig:gas_projection_of_mergers}
\end{figure*}

\begin{table}
\caption{The name and timing of our four merger stages.}
\centering
\begin{tabular}{c ||c c}
\hline \hline
Stage & Starting  & Ending  \\
\hline
Infall & $t_{\mathrm{start}}$ &  $(t_{\mathrm{start}} + t_{\mathrm{max}})/2$  \\
First passage& $(t_{\mathrm{start}} + t_{\mathrm{max}})/2$ &  $t_{\mathrm{max}}$ \\
Coalescence & $t_{\mathrm{max}}$ &  $t_{\mathrm{cls}}$  \\
Post-coalescence &  $t_{\mathrm{cls}}$ &  -  \\
\hline    
\end{tabular}
\label{tab:merger_stages}
\end{table}

Throughout the remainder of this work, we designate the more massive halo/galaxy in a galaxy merger as the primary and the less massive as the secondary. The primary galaxy is also the most massive galaxy or the main galaxy in the \texttt{CosmoRun} simulations. 

Similar to the method outlined in Paper~IX - Part~1, we identified four key moments of our \textit{target} merger: $t_\text{start}$ (the time when the two convex hulls of the merging halos overlap for the first time), $t_\text{fp}$ (the time of the first periapsis), $t_\text{max}$ (the time of the first apoapsis), and $t_\text{cls}$ (the time of coalesence). The timestep preceding $t_\text{start}$ is also referred to as the pre-infall timestep ($t_\text{pre-infall}$). The coalescence criterion, as detailed in Paper~IX - Part~1, is determined using the primary galaxy's radius, the distance between the two progenitor galaxies, their relative velocity, and the velocity dispersion of each galaxy's stellar core. We define a galaxy's stellar core as the 10\% most bound star particles selected at $t_\text{start}$, and we tracked these particle subsets throughout the merging interaction to obtain the distance, relative velocity, and velocity dispersion of the progenitor galaxies. From the timings, we divided the merging interaction into four stages, which are shown in Table~\ref{tab:merger_stages}. The merger's DM mass ratio, stellar mass ratio, and total mass ratio are calculated at $t_\text{pre-infall}$. Table~\ref{tab:merger_params} displays the timings and the mass ratios of the \textit{target merger}.  The reader can refer to Paper~IX - Part~1 for details of other merger parameters, such as the orbital trajectory, the orbital orientation, the gas mass, and the DM mass of the progenitor galaxies. 

Another important timestep for our analysis is the equivalent timestep. Because of the timing discrepancy and variations in merger durations (as shown in fig.~3 of Paper~IX - Part~1), it is essential to select an appropriate timestep for analysis to ensure a fair comparison of the merger outcomes across different codes. We refer to this timestep as the equivalent timestep ($t_{\mathrm{post/eq}}$). It must satisfy two conditions. Firstly, it must have the same elapsed time after the first periapsis to allow the \textit{target} merger in each code to dynamically evolve over the same duration. Secondly, the selected time must fall within the post-coalescence stage (or post-merger stage) yet not too far from the coalescence timestep, when the merger remnant becomes dynamically stable but does not get largely affected by other processes. The only exception in our choice of $t_{\mathrm{post/eq}}$ is with the GEAR simulation, in which the two progenitor galaxies take a significantly longer time to coalesce compared to the other \texttt{CosmoRun} simulations (Table~\ref{tab:merger_params}). By the time GEAR's coalescence is achieved, the main galaxy has already started the infall stage of another merger with a DM mass ratio of $\approx 0.2$. Thus, $t_{\mathrm{post/eq}}$ cannot be selected to be in the post-coalescence stage of GEAR, as the galaxy properties would be contaminated by the effect of the second merger or other processes in between. As a result, for the \textit{target} merger, we chose $t_{\mathrm{post/eq}}$ to be in the post-coalescence stage of all codes but GEAR, for which it falls within GEAR's coalescence stage. Specifically, we chose the equivalent timestep to be~600~Myr after the first periapsis. Despite the uncoalesced nuclei, the galaxy system of GEAR at $t_{\mathrm{post/eq}}$ remains relaxed as a single system with no prominent tidal features at its outskirts. In Section~\ref{sec:effect_on_morphology}, we compare the stellar radial mass distribution and the disc morphology after the merger at $t_{\mathrm{post/eq}}$.

Fig.~\ref{fig:gas_projection_of_mergers} displays the gas projections centred on the main galaxy throughout the merging interaction, showing the system at the key timesteps defined above. For each code, we centre the projections on the primary galaxy and align their normal vector with the system's angular momentum axis at $t_{\mathrm{post/eq}}$ (face-on at $t_{\mathrm{post/eq}}$). By the time of coalescence (fifth column), the systems in all codes have relaxed into a single, coherent structure with only one nucleus.

\begin{table*}
\vspace*{1mm}
\caption{The times of four key moments of the \textit{target} merger, the primary galaxy's stellar mass before the infall, the merger's stellar mass ratio, and the merger's DM mass ratio for the nine \texttt{CosmoRun} codes. The times are listed in Gyr since the Big Bang with the corresponding redshifts. We define $t_\text{start}$ as the start of the merger, when the two DM halos first overlap; $t_\text{fp}$ as the time of first periapsis; $t_\text{max}$ as the time of first apoapsis; and $t_\text{cls}$ as the time of coalescence.}
\centering
\begin{tabular}{c || c | c | c | c | c | c | c}
\hline\hline 
Code & $t_\text{start}(\text{Gyr})|z_\text{start}$  & $t_\text{fp}(\text{Gyr})|z_\text{fp}$ & $t_\text{max}(\text{Gyr})|z_\text{max}$ & $t_\text{cls}(\text{Gyr})|z_\text{cls}$ & $\log_{10}(M_{\bigstar\text{,pri}}/\msun)$  & $\mu_\bigstar$ &  $\mu_\text{DM}$\\ 
\hline
{\sc ART-I} & 1.08 | 5.47 & 1.45 | 4.31 & 1.53 | 4.14 & 1.93 | 3.39 & 8.78 & 0.81 & 0.77   \\
{\sc ENZO} & 1.13 | 5.27 & 1.39 | 4.45 & 1.47 | 4.27 & 1.70 | 3.77 & 8.37 & 0.53 & 0.47    \\ 
{\sc RAMSES} & 1.06 | 5.53 & 1.40 | 4.45 & 1.48 | 4.22 & 1.74 | 3.69 & 8.55 & 0.93 & 0.73    \\
{\sc CHANGA} & 1.10 | 5.39 & 1.42 | 4.37 & 1.49 | 4.19 & 1.85 | 3.50 & 8.33 & 0.68 & 0.60  \\
{\sc GADGET3} & 1.05 | 5.59 & 1.42 | 4.36 & 1.48 | 4.23 & 1.66 | 3.85 & 8.85 & 0.15 & 0.88  \\
{\sc GADGET4} & 1.05 | 5.60 & 1.40 | 4.40 & 1.47 | 4.30 & 1.86 | 3.50 & 8.88 & 0.06 & 0.93  \\
{\sc GEAR} & 1.17 | 5.12 & 1.48 | 4.26 & 1.54 | 4.12 & 2.43 | 2.75 & 8.82 & 0.44 & 0.73  \\
{\sc AREPO} & 1.07 | 5.49 & 1.43 | 4.35 & 1.50 | 4.19 & 1.70 | 3.78 & 8.17 & 0.47 & 0.87  \\
{\sc GIZMO} & 1.08 | 5.48 & 1.44 | 4.36 & 1.49 | 4.23 & 1.88 | 3.47 & 8.08 & 0.50 & 0.73  \\
\hline 
\end{tabular}
\label{tab:merger_params}
\vspace*{2mm}
\end{table*}

\subsection{Defining Galactic Centre}
\label{subsect:galactic_center}

\textsc{HASKAP PIE} defines the centre of a DM halo as the centre of gravity of all bound DM particles to that halo. The DM halo's centre, however, does not always coincide with the centre of mass of the galaxy's baryonic component (gas plus stars), especially during halo interactions. As this study mainly focuses on the changes in the stellar and gas components, it is more appropriate to compute and use the baryonic centre of mass in our analysis. We will refer to this centre as the galactic centre. The galactic centre is identified by an iterative procedure in which we begin by choosing an initial centre, then we iteratively expand out to a larger radius and recalculate the centre of mass until the combined density of gas and stars exceeds a specified overdensity threshold relative to the universe's critical density at a given redshift. We find that an overdensity threshold of 2000 works well for this method. The initial centre can be selected as the centre of mass of the galaxy's stellar core, or the centre of mass of the inner-radius stars for a more general case, or the galactic centre found in a previous snapshot. As remarked by Paper~\citetalias{Jung+2025a}, this method works well for galaxies with an extended mass or a non-unimodal distribution, particularly in the case of RAMSES and GEAR.

\section{Results}
\label{sec:effect_on_morphology}

In Paper~IX - Part~1, we identified three patterns of how the star formation rate (SFR) evolves during the merging interaction. These patterns depend on the type of stellar feedback that each code group employs. First, codes that use both kinetic feedback and thermal feedback (ART-I, GADGET-3, GADGET-4, and GIZMO) display a starburst starting from the beginning of the first passage stage and persisting until the first half of the coalescence stage, after which the SFR declines rapidly. Second, codes that use thermal feedback without kinetic feedback (ENZO, RAMSES, CHANGA, GEAR, and AREPO-T) show a general growth of SFR even after coalescence with no significant declines. Third, codes that implement the delayed cooling (RAMSES, GADGET-3, and GEAR) or radiation pressure (ART-I) prescription experience an SFR evolution with more short-timescale, small-amplitude fluctuations. Moreover, the delayed cooling and superbubble schemes inhibit gas from reaching high central density. In this study, we will show that the type of stellar feedback not only impacts the star formation response but also influences the stellar morphology of the merger remnant. We will also discuss how the morphological transformation relates to the SFR evolution trend. 

\subsection{Stellar radial distribution}
\label{subsect:radial_distribution}

\begin{figure*}[tbh]
    \centering
    \includegraphics[width=\linewidth]{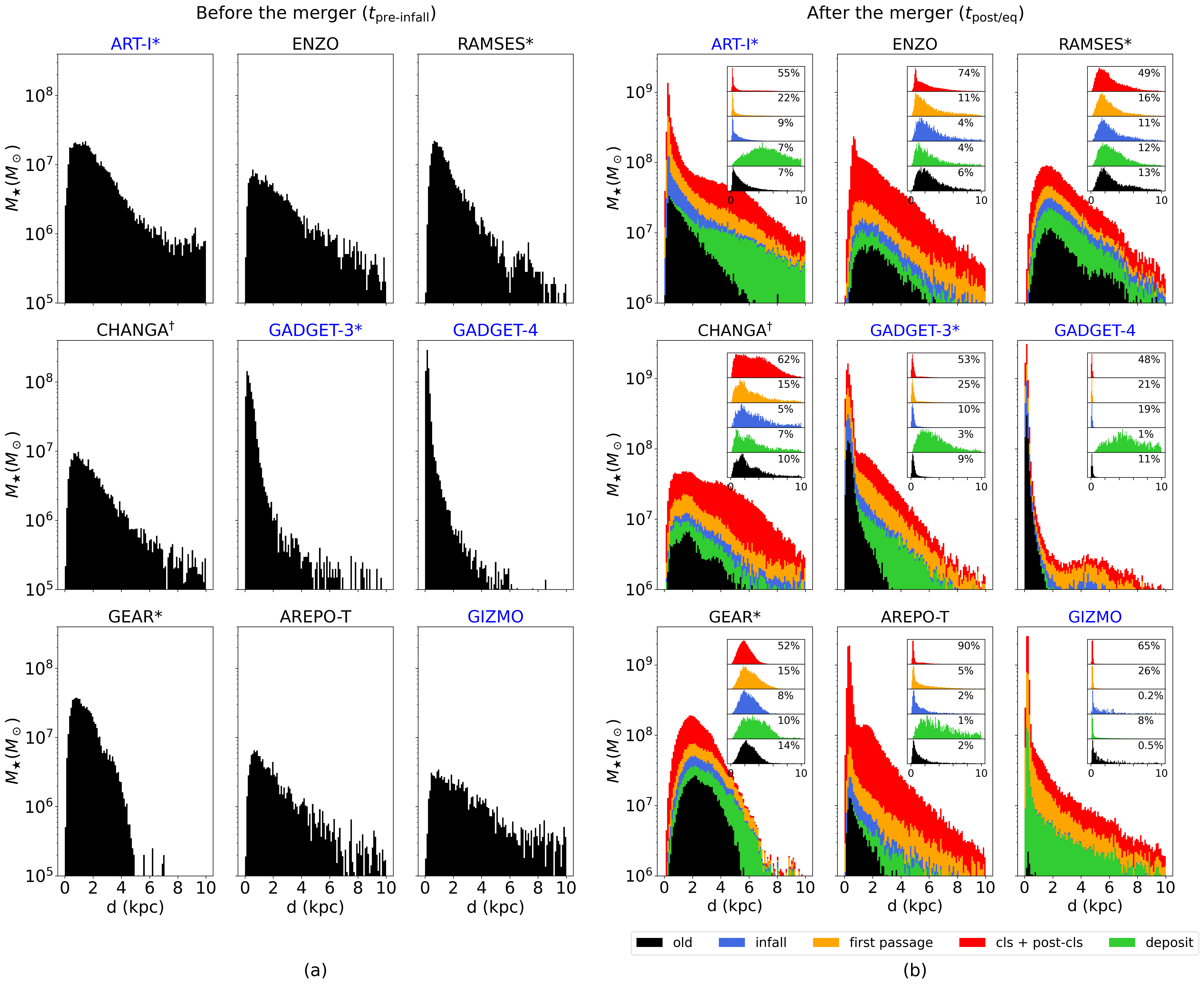}
    \caption{The radial mass distribution of the main galaxy across the codes before and after the \textit{target} merger, shown in panels~(a) and (b), respectively. For better visualization, the y-limit is set differently between panel~(a) and panel~(b). In panel~(b), the total stellar distribution is decomposed into five groups: the old stars existing in the main galaxy before the merger ("old"), stars formed during the infall stage ("infall"), the first passage stage ("first passage"), the coalescence plus post-coalescence stage ("cls + post-cls"), and stars deposited from the secondary galaxy ("deposit"). We note that because the y-axis is logarithmic, the segment heights are not linearly additive. The insets in the top right corner of each subplot in panel~(b) display the mass profile of each of these stellar groups. Unlike the main subplots, these insets use a linear scale and have different ranges on their y-axes to emphasise their distribution's shape. The fractional mass of each stellar group relative to the galaxy’s total stellar mass is also reported in each inset. The formatting of the code names follows that of Fig.~\ref{fig:gas_projection_of_mergers}. Different feedback prescriptions produce distinct morphological changes in the stellar mass profile.}
    \label{fig:radial_star_distribution}
\end{figure*}

We first investigate where stars formed during different merger stages are located in the merger remnant. Fig.~\ref{fig:radial_star_distribution} displays the radial stellar mass distribution of the main galaxy, with panel~(a) showing the pre-merger state at $t_{\mathrm{pre-infall}}$ and panel~(b) showing the post-merger state at $t_{\mathrm{post/eq}}$. The post-merger distribution is further decomposed into five stellar groups: the main galaxy's pre-existing stars prior to the merger, stars formed during the infall stage, stars formed during the first passage stage, stars formed during the coalescence stage and/or up to the equivalent timestep, and stars deposited from the secondary galaxy. We define deposit stars (in green) as stars that are, at any given time during the interaction, more gravitationally bound to the secondary galaxy than to the primary galaxy. In other words, the deposit stellar group includes all stars existing in the secondary galaxy before the infall, as well as stars formed during the interaction that are more bound to the secondary galaxy (note that a star can be bound to both the main galaxy and secondary galaxy during a merger). Because of tidal stripping, these deposit stars have an extended radial distribution after accreting onto the main galaxy, as shown in the green inset histograms in Fig.~\ref{fig:radial_star_distribution} for most of the codes. 

At the beginning of the merger, except for GADGET-3 and GADGET-4, the participating codes agree relatively well on the stellar mass profile of the main galaxy, as shown by panel~(a) of Fig.~\ref{fig:radial_star_distribution}. Nonetheless, the profile shows considerable divergence after the merger. The main galaxy in ART-I, ENZO, AREPO-T, and GIZMO becomes more compact after interaction, whereas the galaxy in RAMSES, CHANGA, and GEAR becomes more extended. This compaction can also be seen in the gas projection plots in the sixth column of Fig.~\ref{fig:gas_projection_of_mergers}. In GADGET-3 and GADGET-4, the main galaxy already undergoes an early compaction before the \textit{target} merger, which is caused by a preceding strong starburst at $z \approx 6$ (see Appendix B of Paper~IX - Part~1 for a detailed analysis). Indeed, the stellar half-mass radii of GADGET-3 and GADGET-4 (329 pc and 168 pc, respectively) are about an order of magnitude smaller than the average value of the other codes. As these half-mass radii are comparable to the spline softening scale, $\epsilon_{\rm sp} = 2.8\,\epsilon \approx 224~{\rm pc}$ (\citealp{Springel+2005}, for a Plummer softening length of $\epsilon = 80~{\rm pc}$ set in the \texttt{CosmoRun} suite), the pre-merger galaxy sizes of GADGET-3 and GADGET-4 are already near the spatial resolution limit. Because the gravitational force is progressively suppressed below $\epsilon_{\rm sp}$ and hence the gravity that drives dissipative contraction is artificially weak, it is harder for the galaxies to contract to smaller radii. As softening inhibits any further compaction, the half-mass radii change very little throughout the whole merging interaction, measuring $334~{\rm pc}$ and $133~{\rm pc}$ at $t_\text{post/eq}$ for GADGET-3 and GADGET-4, respectively. We also want to note that though the two galaxies are both compact, the GADGET-3 galaxy is slightly more extended, likely because of its delayed cooling feedback. As shown in Paper~IX - Part~1, delayed cooling feedback can create a spatially broader gas distribution.

For the other seven codes, the connections between star formation response and stellar feedback types identified in Paper~IX - Part~1 manifest as three distinct morphological changes in the radial stellar mass distribution: rapid compactification, late compactification, and expansion. Hereafter, we refer to them as Group 1, Group 2, and Group 3, respectively. 

\textbf{Group 1 - Rapid compactification}: Codes that employ kinetic feedback produce a more compact merger remnant (ART-I and GIZMO, denoted in blue; GADGET-3 and GADGET-4 cannot be used to test this behaviour due to their galaxy's size being already at the numerical floor, as mentioned above). The new stars formed throughout the interaction are formed primarily in the galaxy's inner region (at radii < 1 kpc), as seen across the blue (infall stage), yellow (first passage stage), and red (coalescence plus post-coalescence stages) histograms in the insets in panel~(b) of Fig.~\ref{fig:radial_star_distribution}. These three histograms also have approximately similar shapes and are much more compact than the old stellar population (black histograms in the insets), suggesting that the compaction starts as soon as the end of the infall stage for these codes. The compaction is caused by gas losing angular momentum and falling into the galactic centre, a process often referred to as wet compaction, as discussed in Paper~\citetalias{Jung+2025a}. The early, rapid compaction is consistent with the observation that the \textit{target} merger in these codes undergoes an earlier starburst relative to other \texttt{CosmoRun} codes. Due to an early pronounced starburst, ART-I, GADGET-3, GADGET-4, and GIZMO also have the largest percentage of stars formed during the first passage stage ($> 20\%$) in the merger remnant evaluated at $t_{\mathrm{post/eq}}$. 
    
\textbf{Group 2 - Late compactification}: Codes with only thermal feedback (ENZO and AREPO-T, denoted in black with no additional annotations) also exhibit compaction after the \textit{target} merger, yet this compaction happens mainly later in the interaction. We note that, because of the logarithmic scaling, the high stellar density in the central region of ENZO appears less prominent. Compared to the earlier stages in the interaction, the radial distribution of the stars formed during the coalescence and post-coalescence stages is much more concentrated in the galactic central region. Indeed, throughout the interaction, the distribution stays relatively extended and comparable between the infall stage (blue histogram) and the first passage stage (yellow histogram), and then becomes more compact during the coalescence and post-coalescence stage (red histogram). In conjunction with the continuously increasing SFR identified in codes that use solely thermal feedback (Paper~IX - Part~1), this finding implies that the collapse of gas into a central starburst is more pronounced at a later merger stage. Using thermal feedback alone suppresses the merger-driven inflow in the first periapsis, and gas can only cool down and collapse towards the galactic centre when the second periapsis and final coalescence happen. In other words, the compaction of ENZO and AREPO-T occurs primarily when the two galaxies coalesce, rather than in the early stages of the interaction as in codes employing kinetic feedback in their models. Indeed, the merger remnants of ENZO and AREPO-T have the largest percentage of stars formed during the coalescence and post-coalescence stages (74\% and 90\%), considerably larger than that value of other codes ($\approx 50\text{--}60\%$). This large mass percentage is even more notable considering that the "cls + post-cls" interval spans only slightly more than half the interval between $t_\mathrm{start}$ and $t_\mathrm{post/eq}$, demonstrating the importance of the coalescence and post-coalescence stages for star formation and compaction in these two codes. 
    
\textbf{Group 3 - Expansion}: For codes that do not utilise kinetic feedback but use the delayed cooling scheme (RAMSES and GEAR, denoted in black with an asterisk sign $*$) or the superbubble scheme (CHANGA, denoted in black with a dagger sign $\dagger$), the merger remnant becomes more extended and increases in size compared to the galaxy at $t_{\mathrm{pre-infall}}$. For RAMSES and GEAR, the radial stellar distribution remains approximately the same for stars formed across different merger stages and even for the older population that existed before the merger, as shown by the black, blue, yellow, and red inset histograms in panel~(b) of Fig~\ref{fig:radial_star_distribution}. These two codes also have the largest percentage of old stars (black histogram) in the merger remnant at $t_{\mathrm{post/eq}}$. As previously explained, the delayed cooling scheme hinders gas from dissipatively collapsing to the centre despite the dynamical disturbance of the \textit{target} merger, leading to a broader distribution of gas (also shown in Fig.~\ref{fig:gas_projection_of_mergers}). As RAMSES and GEAR only impose a density threshold and not a temperature threshold for star formation, stars can still form in warm-hot dense gas, resulting in an extended stellar distribution. For CHANGA, its superbubble feedback prescription drives strong outflows that expel gas into the circumgalactic medium (\citealp{Keller+2014}, Papers~\citetalias{Jung+2025a} and~IX - Part~1). This preventative feedback suppresses the cooling and inflow of gas toward the galactic centre, thereby inhibiting the dissipative contraction that would otherwise drive a compaction event \citep{Dekel+2014}. In CHANGA, the stars formed during the coalescence/post-coalescence stage (red histogram) are more broadly distributed than the stars formed during the infall and the first passage stages (blue and yellow histograms). In other words, the outskirts of the merger remnant are dominated by young stars, and this young population makes the galaxy grow in size. This size growth results from CHANGA's strong feedback, which pressurizes and ejects gas from the galaxy to the circumgalactic medium (CGM) to yield a more extended gas distribution. The recycled gas returning from the CGM can also fuel star formation at larger radii (see fig.~6 and Appendix~C of Paper~IX – Part~1). It is important to note that, as indicated in Paper~IX - Part~1, CHANGA experiences a delayed starburst around~500~Myr after coalescence when gas re-accretes back onto the main galaxy. After that delayed starburst, the main galaxy eventually undergoes compaction. 

\subsection{Relationship between merger-driven burst fractions and galaxy compaction}
\label{subsect:burstfraction_vs_compaction}

\begin{figure}[tbh]
    \centering
    \includegraphics[width=\columnwidth]{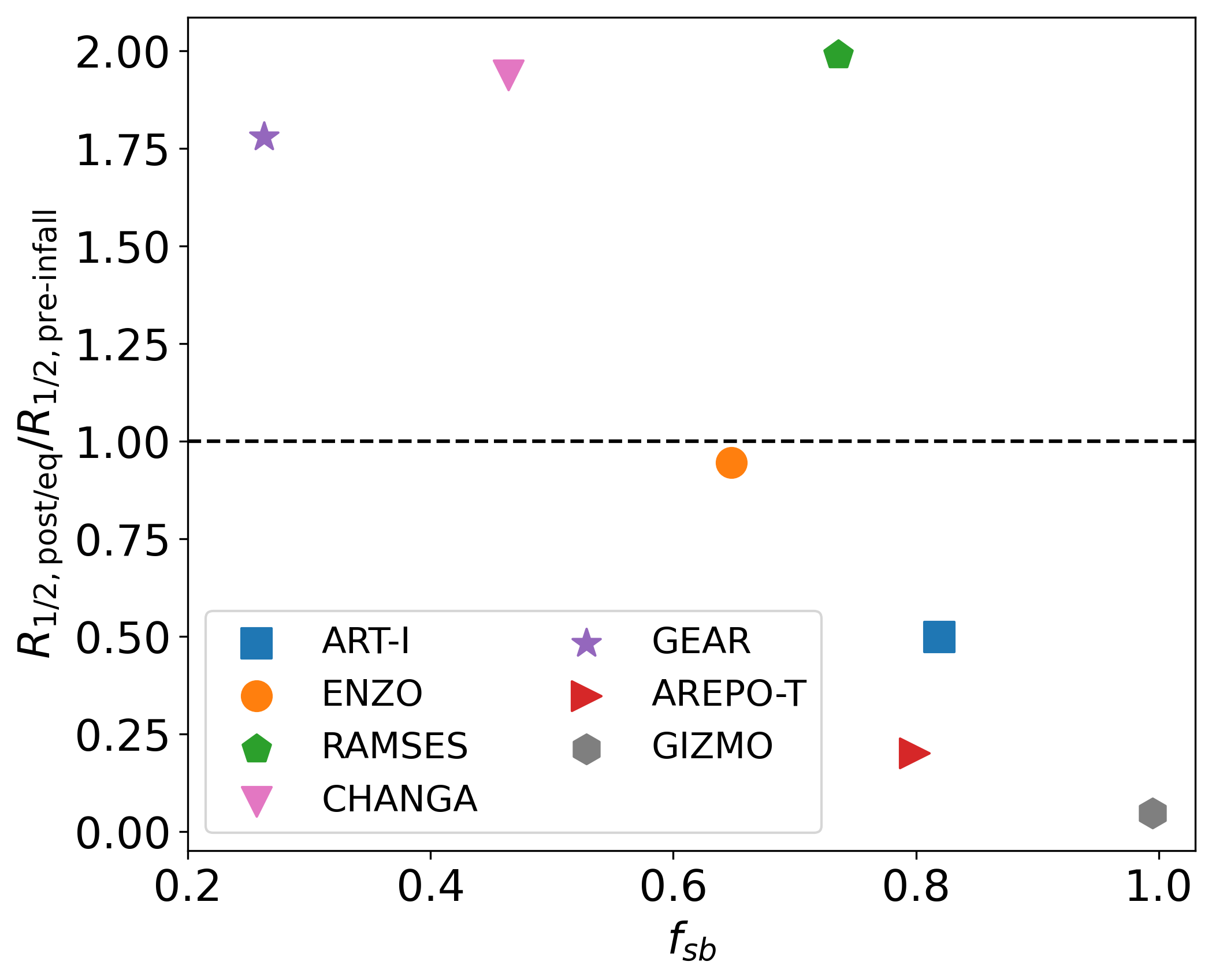}
    \caption{The ratio between the post-merger and pre-merger half-mass radius as a function of the burst fraction of the \textit{target} merger ($f_\text{sb}$). For codes without delayed cooling feedback, a stronger merger-driven starburst corresponds to a more compact galaxy after the merger. GADGET-3 and GADGET-4 are omitted from the plot because their burst fraction cannot be reliably computed.}
    \label{fig:HalfMassRadiusRatio_vs_burstFraction}
\end{figure}

We proceed to explore how the changes in the galaxy's size correspond to the merger-driven star formation enhancement. To quantify this enhancement, we adopted a new method that does not rely on a control, isolated galaxy sample. This approach is particularly advantageous for cosmological zoom-in simulations, where it is not easy to re-simulate the galaxy without a specific merger or assemble a large sample of isolated, similar-mass galaxies within the same simulation box. We present the method in full in Paper IX - Part~1 and provide a brief summary as follows.
We assume that the sSFR of a galaxy remains constant when the galaxy experiences only smooth accretion and no mergers. Because there are no significant merger events occurring within~100 Myr before the \textit{target} merger \footnote{GADGET-3 and GADGET-4 are excluded from the burst fraction calculation because they experience a strong starburst shortly before the \textit{target merger}. This strong starburst makes the determination of the baseline sSFR unrepresentative of the main galaxy in isolation.}, we chose the lowest sSFR within this period as the baseline sSFR. Then, we calculated the baseline stellar mass of the main galaxy if the \textit{target} merger did not happen. With the baseline and the actual stellar masses from the simulations, we define the burst fraction as the excess of the actual stellar mass over the baseline, normalised by the actual mass, $f_\text{sb} = (M_\text{actual} - M_\text{baseline})/M_\text{actual}$. Fig.~\ref{fig:HalfMassRadiusRatio_vs_burstFraction} shows the fractional change of the main galaxy's stellar half-mass radius before and after the merger as a function of the burst fraction. 

With the exception of RAMSES and GEAR, the two codes implemented with delayed cooling feedback, we find a negative correlation between $R_{1/2,\text{post/eq}}/R_{1/2,\text{init}}$ and $f_{sb}$. The galaxies that experience a decrease in size have a burst fraction larger than~0.6, whereas the galaxies that increase in size have a smaller burst fraction. Codes with strong preventative feedback, such as CHANGA, suppress the collapse of gas to the galactic center, resulting in a more extended distribution of gas and stars. This suppression also causes the burst fraction to be small because gas cannot cool down easily. On the other hand, for codes with more efficient dissipative gas collapse (ART-I, ENZO, AREPO-T, and GIZMO), gas can reach a higher central density, leading to a more intense starburst and a more pronounced compaction. In summary, mergers with higher burst fractions are associated with more compact remnants, whereas lower burst fractions are associated with more extended remnants. This relationship is also hinted at by the semi-analytic model of \cite{Covington+2008}. In their model, a larger dissipative strength of a merger ($f_{k}$) corresponds to a higher fraction of new stars ($f_\text{new}$) and a smaller stellar-half-mass radius of a remnant, though their relationship is not necessarily linear. It is important to note that the relation between $R_{1/2,\mathrm{post/eq}}/R_{1/2,\mathrm{init}}$ and $f_{\mathrm{sb}}$ is correlative, not causal. 

For RAMSES and GEAR, the two outliers to the correlation (especially RAMSES), we attribute these exceptions to the codes' delayed-cooling schemes, which suppress radiative losses for an extended period (10~Myr in RAMSES and 5~Myr in GEAR). Furthermore, because we decided not to impose any temperature threshold for star formation in \texttt{CosmoRun}, stars can still form in warm--hot dense gas. The delayed cooling scheme prevents dissipation of thermal energy through numerically turning off gas cooling, yet the absence of the temperature criterion still permits the burst to proceed in the undissipated gas. Indeed, Paper~\citetalias{Jung+2025a} showed that the main galaxy in RAMSES and GEAR does not undergo a compaction event even at $z = 2$, when the galaxies in most codes become compact. In short, we hypothesise that RAMSES and GEAR do not strictly follow the correlation because stars can still form in highly undissipated hot gas, which leads to a broader stellar distribution. 
Therefore, the correlation between $R_\text{1/2,post/eq}/R_\text{1/2,pre-infall}$ and $f_\text{sb}$ holds across codes, but only under specific numerical implementations.

Even without the two outlier cases, we caution that this relationship still has some caveats to be fully established. The reason is that \texttt{CosmoRun} is constructed for comparison of the same merger between different codes, rather than comparison of different mergers in a single code. The codes' distinct feedback schemes and numerical architectures may impose a systematic offset in the merger remnant's compactness that is unrelated to the burst fraction. Moreover, the correlation is defined by a small number of points (five) in Fig.~\ref{fig:HalfMassRadiusRatio_vs_burstFraction}. Hence, confirming a genuine relation between $R_\text{1/2,post/eq}/R_\text{1/2,pre-infall}$ and $f_\text{sb}$ would require a large suite of mergers with varying star formation enhancement evolved within each code individually, which is beyond the scope of \texttt{CosmoRun}.

\subsection{Disc formation after the merger}
\label{subsect:disk_formation}

\begin{figure*}[tbh]
    \centering
    \includegraphics[width=\linewidth]{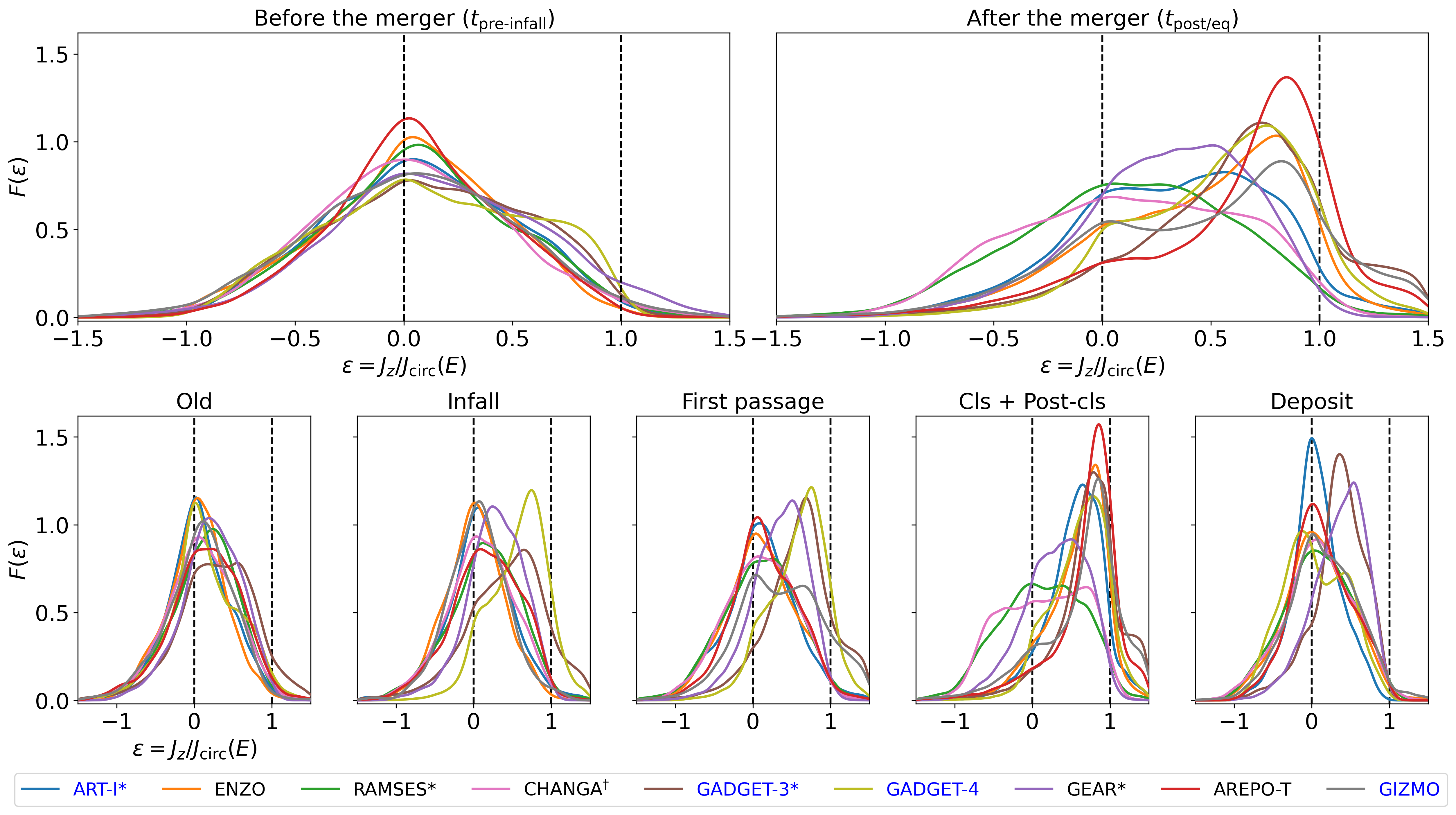}
    \caption{The mass-weighted distribution of orbital circularity $\epsilon$ for star particles in the main galaxy. Top row: the distribution at the pre-infall timestep (left) and at the equivalent timestep (right). Bottom row: the $\epsilon$ distribution of stars in the five stellar groups in the main galaxy after the \textit{target} merger: old stars existing in the main galaxy before the merger, stars formed during the merger stages listed in Table~\ref{tab:merger_stages}, and stars accreted from the secondary galaxy. The stellar groups and the formatting of the code names are similar to those in Fig.~\ref{fig:radial_star_distribution}. As with the impact on the stellar radial profile, different feedback types produce distinct responses in stellar disc formation after the \textit{target} merger.}
    \label{fig:circularity_distribution}
\end{figure*}

\begin{figure*}[tbh]
    \centering
    \includegraphics[width=\linewidth]{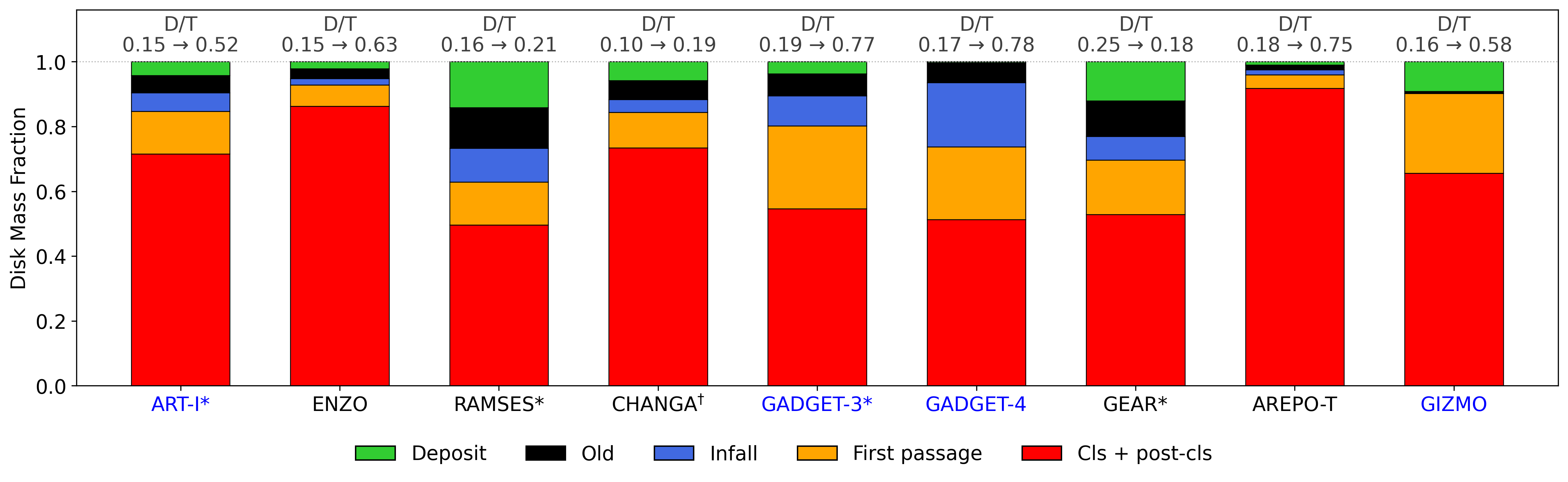}
    \caption{The stacked mass fraction contributed by each merger stage to the merger remnant's disc, evaluated at $t_\mathrm{post/eq}$. The disc-to-total ratios (D/T) of the main galaxy at $t_\text{pre-infall}$ (left) and at $t_\mathrm{post/eq}$ (right) are reported above the bar of each code. Except for RAMSES, CHANGA, and GEAR, all codes build up a substantial stellar disc after the interaction (D/T~>~0.5). In codes with only thermal feedback (ENZO and AREPO-T), the disc is composed predominantly of stars formed during the coalescence and post-coalescence stage. In codes adopting kinetic feedback along with thermal feedback (ART-I, GADGET-3, GADGET-4, and GIZMO), the disc also incorporates a considerable number of stars formed during earlier merger stages.}
    \label{fig:disk_decomposition}
\end{figure*}

\begin{figure*}[tbh]
    \centering
    \includegraphics[width=\linewidth]{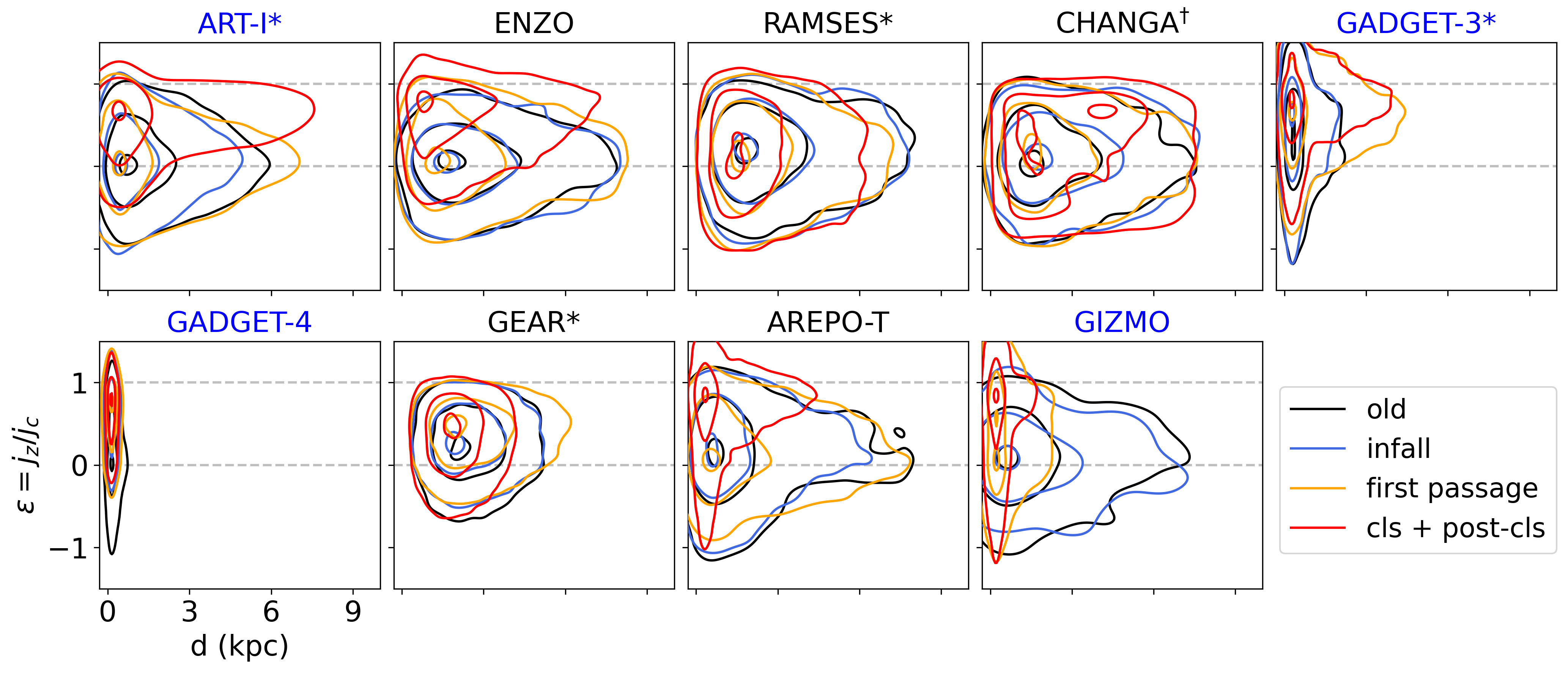}
    \caption{The distribution of stars formed during different merger stages on the radial distance-orbital circularity plane, evaluated at $t_\mathrm{post/eq}$. A kernel density estimate is applied to draw the 10\%, 50\%, and 95\% contour levels of the distributions. For better visualization, only stars within $0.2R_{200c}$ are used for the kernel density estimate. These plots combine the results from Figs.~\ref{fig:radial_star_distribution} and ~\ref{fig:circularity_distribution}, offering us a comprehensive picture of the morphological changes (compaction and disc formation) of the main galaxy after the \textit{target} merger. Codes using kinetic feedback (blue names) show a gradual sign of compaction and disc formation throughout the merging event; codes using only thermal feedback (black names with no additional annotation) become more compact and discy primarily later in the coalescence stage; and codes using thermal feedback only with delayed cooling or superbubble schemes (black names with an asterisk $*$ or a dagger $\dagger$ sign) do not show significant signs of compaction or disc formation after the merger.}
    \label{fig:circularity_vs_distanc}
\end{figure*}

As discussed in Paper~\citetalias{Jung+2025a}, in many of the codes, the \textit{target} merger induces the main galaxy to form a stellar disc. Here, we dive deeper into the investigation by dissecting the disc structure into components that formed during different merger stages. To characterise the disc formation, we use the stellar kinematic properties and define the orbital circularity $\epsilon = j_{z}/j_\text{circ}(E)$ parameter, where $j_{z}$ is the z-component specific angular momentum of a star particle around the galactic center and $j_\text{circ}(E)$ is the maximum angular momentum at a given orbital energy \citep{Abadi+2003}. For a dispersion-dominated galaxy, the distribution of $\epsilon$ peaks around zero. For a rotation-dominated galaxy, the $\epsilon$ distribution peaks near one. Thus, a disc galaxy with a bulge typically exhibits a bimodal $\epsilon$ distribution, with one peak near $\epsilon \approx 0$ and the other near $\epsilon \approx 1$. A star has $\epsilon < 0$ if it orbits in a counter-direction with the galaxy's total angular momentum. As we calculated the $j_\text{circ}(E)$ profile using the four-point azimuthal average on the disc plane for various radii and interpolated between them \citep{Liang+2025}, $\epsilon$ can also be slightly larger than~$1$ or smaller than~$-1$ if a star's 3D potential mismatches with the interpolated midplane $j_\text{circ}(E)$. Thus, we placed a limit of $|\epsilon| < 1.5$ in our analysis, though star particles with extreme values of circularity are typically rare.

Similar to Paper~\citetalias{Jung+2025a}, we followed the procedure of \cite{Liang+2025} to calculate $\epsilon$ and decompose the morphological components, except that we included all stars within the halo's convex hull instead of within $5\times R_{1/2}$ as in the method paper. \cite{Liang+2025}'s method can decompose a galaxy into four components: bulge, halo, thin disc, and thick disc. Nevertheless, we chose to combine the bulge and the halo component into a single "spheroid" component, as the decomposition is less straightforward for some codes at high redshift. Since the spheroid is assumed to possess no net rotation, we modelled it with the $\epsilon$ distributions that are symmetric about 0. To assign star particles to the spheroid component, we took all particles with $\epsilon < 0$ and added a mirrored population drawn from the particles with $\epsilon > 0$ via Monte Carlo sampling, such that the sampled population follows the reflection of the $\epsilon < 0$ distribution. The remaining star particles are assigned to the disc components. The disc-to-total (D/T) value is calculated as the mass of both the thin disc and thick disc components divided by the total stellar mass. One exception to this approach is the GEAR galaxy. Compared to other codes, GEAR takes a significantly longer time for the \textit{target} merger to reach coalescence (fig.~2 of Paper~IX - Part~1). This prolonged orbital phase complicates the kinematic decomposition because the two progenitors' intrinsic bugles orbit around the centre of mass of the galaxy system, thus displacing the peak of the system's circularity distribution from zero. In other words, the total angular momentum of the system is dominated by the orbital angular momentum of the two stellar cores instead of the main galaxy's intrinsic rotational angular momentum. As shown in Fig.~\ref{fig:circularity_distribution}, for GEAR, the peaks of the stellar circularity distributions of all five stellar groups are systematically offset toward 0.2. Because of this, we adopted the decomposition method by \cite{Kannan+2015}, which does not assume the circularity distribution of the spheroid component to be symmetric around zero. We provide further details in Appendix~\ref{subsect:GEAR_Disk_Decomposition}.

As shown in the top left subplot in Fig.~\ref{fig:circularity_distribution}, initially for all codes, the highest peak of the circularity distribution is located around zero, indicating that the main galaxy is more dispersion-dominated before the \textit{target} merger. Indeed, as shown in  Fig.~\ref{fig:disk_decomposition}, all codes exhibit relatively small D/T fractions ($< 0.3$) at $t_\text{pre-infall}$. Moreover, most codes display only a unimodal $\epsilon$ distribution centred at $\epsilon \approx 0$, which implies that their pre-merger main galaxies are spheroidal. Nonetheless, GADGET-4 is an exception. Its $\epsilon$ distribution has a distinct local maximum at $\epsilon \approx 0.8$ despite the galaxy having a low D/T fraction. We emphasize that these two features are not contradictory. In our method, the D/T fraction is set not by the position or amplitude of the $\epsilon$ peak associated with the disk component, but only by the amount of star particles with $\epsilon < 0$ ($\text{D/T} \approx 1-2M_{\bigstar,\epsilon<0}/M_{\bigstar,\text{total}}$). Therefore, a high-amplitude $\epsilon$ peak near one can still imply a kinematically distinct stellar disc, even though that disc may be relatively low-mass compared to the total stellar mass. Even though less pronounced than GADGET-4, GADGET-3 likewise shows a similar indication of a distinct disc structure pre-merger. The disc formation of GADGET-3 and GADGET-4 before the \textit{target} merger is attributed to the starburst event at $z \approx 6$, which is discussed in more detail in Paper~IX - Part~1.

From Figs.~\ref{fig:circularity_distribution} and \ref{fig:disk_decomposition}, we see that after the \textit{target} merger, the merger remnants in ART-I, ENZO, GADGET-3, GADGET-4, ARPEO, and GIZMO develop a rotational structure with younger stars formed during the interaction being the main component. The D/T values for these codes all exceed~0.5 at the equivalent timestep, showing that the remnants are rotation-dominated. In all codes, the old stellar population in the merger remnant (first column of the bottom row) resides in a spheroidal component dominated by random motion. The same behavior appears even in GADGET-4, whose galaxy has a disc component before the merger. This transition of pre-infall stars from constituting a rotational-support structure to a pressure-support structure is caused by a process called violent relaxation \citep{Lynden-Bell+1967}. During a merger, a rapidly fluctuating gravitational potential deflects star particles from their orbit and smooths out the phase space distribution of particles, leading to a system dominated by random motions. Violent relaxation is common in many simulations of binary mergers between disc galaxies \citep{Hopkins+2009, Kannan+2015}. 

To elucidate the connection between the stellar mass radial profile (Fig.~\ref{fig:radial_star_distribution}) and the bulge/disc components (Fig.~\ref{fig:circularity_distribution}), we show in Fig.~\ref{fig:circularity_vs_distanc} the distribution of star particles at $t_\mathrm{post/eq}$ on the radial distance-orbital circularity plane. Stars are separated into different merger stages, and a kernel density estimate is used to make the 10\%, 50\%, and 95\% contour levels. The levels represent iso-proportions of the density, so that the 95\% contour bounds the highest-density region and corresponds to the innermost line, and the 10\% contour corresponds to the outermost. We omit the "Deposit" group in this figure for a clearer visualization. As with the effect on the stellar radial profile, we also notice three distinct impacts on the merger remnant's disc formation based on the type of the simulation's stellar feedback schemes. The group numbers match those defined in Section~\ref{subsect:radial_distribution}.

\textbf{Group 1 - Early disc formation}: For codes with kinetic feedback (blue names), the main galaxy shows a gradual development of a stellar disc structure from the infall stage to the coalescence stage. In Fig.~\ref{fig:circularity_vs_distanc}, we can observe that the centre of the highest-density contour level (the innermost contour) is gradually shifted up from $\epsilon = 0$ to $1$ from the old stellar population to the coalescence/post-coalescence stellar population (black to blue to yellow to red lines). In other words, the disc starts forming early in the merger event, though stars that are born later in the interaction still contribute more to the disc. According to Fig.~\ref{fig:disk_decomposition}, stars formed during the infall stage (blue portion) and the first passage stage (yellow portion) account for~20\% to~50\% of the stellar disc mass. For GIZMO, Fig.~\ref{fig:circularity_vs_distanc} also displays a conspicuous case of wet compaction, where young stars formed in the merger-driven starburst are much more concentrated at the galactic centre compared to older stars. Furthermore, not only do these stars form at a smaller radius, but they also make up the newly formed stellar disc. 
    
\textbf{Group 2 - Late disc formation}: For codes with only thermal feedback (ENZO and AREPO-T), the formation of the disc mainly starts during the coalescence and post-coalescence stage. Fig.~\ref{fig:disk_decomposition} displays that the coalescence and post-coalescence stages (red portion) account for more than~85\% of the stellar disc mass, whereas the earlier stages (blue and yellow portions) contribute less than~10\%. Indeed, the contribution of earlier merger stages is considerably smaller than that of codes with kinetic feedback (Group~1), highlighting the dominant role of the coalescence and post-coalescence stages in disc formation in ENZO and AREPO-T. Stars formed during the infall stage and the first passage stage are still dominated by random motion, as shown in Fig.~\ref{fig:circularity_distribution} (the peak of the circularity distribution is at zero).  Indeed, Fig.~\ref{fig:circularity_vs_distanc} shows that, on average, stars formed earlier in the interaction more resemble the old population (black line) in the circularity and radial distance space. On the other hand, the stellar population born during and after the coalescence stage is located closer to the galactic centre and is rotation-dominated (the highest-density contour level is further shifted closer to one). As gas gets heated up due to thermal feedback, the gas's increased thermal energy drives
it out of the galaxy. After the two nuclei merge and the central potential relaxes, this gas can cool, fall back onto the galaxy, and form a rotationally supported gas disc and stellar disc. The SFR pattern also reflects this gas behavior, in which ENZO and AREPO-T display the highest SFR after coalescence (Paper XI - Part 1).
    
\textbf{Group 3 - Negligible disc formation}: For codes that utilise only thermal feedback with the delayed cooling scheme (RAMSES and GEAR) or the superbubble scheme (CHANGA), no stellar disc is clearly present in the galaxy at $t_\text{post/eq}$. The D/T values of these codes remain low after the interaction and change very slightly compared to the pre-infall values (Fig.~\ref{fig:disk_decomposition}). The merger remnant remains dominated by random stellar motion. The $\epsilon$ distribution before and after the merger both stay relatively centred at zero. For GEAR, even though the $\epsilon$ distribution peaks around~0.2, the apparent rotational signal arises from the two galactic nuclei being uncoalesced at the equivalent timestep, rather than from the presence of an actual stellar disc (more details in Appendix~\ref{subsect:GEAR_Disk_Decomposition}). Moreover, on the radial distance-circularity space (Fig.~\ref{fig:circularity_vs_distanc}), the distributions of stars from different merger stages are relatively similar to each other, indicating an absence of significant disc formation in the main galaxies in RAMSES, CHANGA, and GEAR after the \textit{target} merger. 
    
This major merger in AGORA \texttt{CosmoRun} also gives us an example of how a major merger can turn an elliptical galaxy into a disc galaxy, which arises in all codes but CHANGA, RAMSES, and GEAR. A common theoretical picture is that mergers disrupt the disc structure of galaxies and turn them into being more elliptical \citep{Hopkins+2008}. Even though a disc can reform after a major merger \citep{Hopkins+2009}, there has been little discussion in the literature of an evolutionary pathway in which an elliptical galaxy becomes discy after a major merger, especially in observations. Using a sample of~38 disc galaxies in the Illustris simulation, \cite{Peschken+2020} also supported our result by finding examples of clumpy ellipsoidal progenitor galaxies quickly forming a disc after a major merger. Similar to this study's findings, they also found that the stellar discs in these merger remnants are made up of young stars, whereas the old stars are more distributed in the spheroid component. They attributed the disc formation to the amount of gas available to the system after the merging event, where wet mergers create disc galaxies and dry mergers create elliptical galaxies. Our results show that the adopted stellar feedback of a simulation also plays an important role in determining the morphological outcome. For instance, although GEAR and CHANGA have considerably more gas than other codes (Paper~IX - Part~1), their disc formation is still suppressed by strong gas heating due to feedback. More observational evidence and a larger sample size of mergers of ellipsoidal galaxies with different feedback types should be explored in future work to understand how common this pathway is.  

\subsection{Angular momentum orientation of the merger remnant's disc}

\begin{figure*}[tbh]
    \centering
    \includegraphics[width=\linewidth]{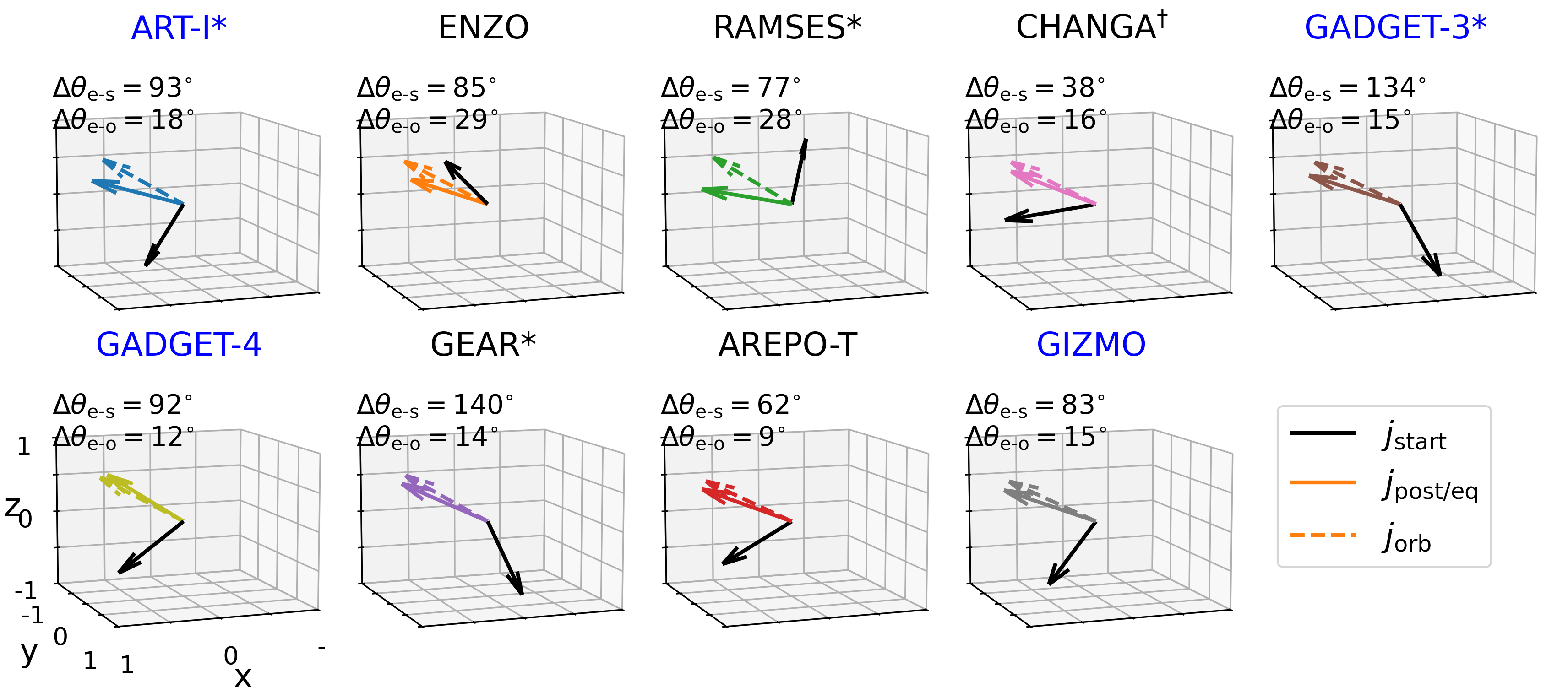}
    \caption{The unit vectors showing the directions of the main galaxy's rotational angular momentum at $t_\text{start}$ ($\vec{j}_\text{start}$, solid black arrows), the merger remnant disc's rotational angular momentum at $t_\text{post/eq}$ ($\vec{j}_\text{post/eq}$, solid coloured arrows), and the orbital angular momentum of the merging interaction evaluated at the first periapsis ($\vec{j}_\text{orb}$, dashed-line coloured arrows). $\vec{j}_\text{post/eq}$ is calculated using only the star particles assigned to the disk component from the kinematic decomposition. $\Delta\theta_\text{e-s}$ is the angle between  $\vec{j}_\text{post/eq}$ and $\vec{j}_\text{start}$, and $\Delta\theta_\text{e-o}$ is the angle between $\vec{j}_\text{post/eq}$ and $\vec{j}_\text{orb}$. The formatting of the code names follows that of Fig.~\ref{fig:radial_star_distribution}. The orientation of the merger remnant's disc is heavily influenced by the orbital angular momentum of the interaction. For codes with a pre-merger disc (GADGET-3 and GADGET-4), the \textit{target} merger destroys that disc, and a new disc reforms quickly with a different orientation.}
    \label{fig:angular_momentum_direction_change}
\end{figure*}

After finding that a stellar disc forms for six out of nine \texttt{CosmoRun} codes after the \textit{target} merger, we examine the orientation of the disc and what influences that orientation. Fig.~\ref{fig:angular_momentum_direction_change} shows the unit vectors of $\vec{j}_\text{start}$ (the main galaxy's rotational angular momentum at $t_\text{start}$), $\vec{j}_\text{post/eq}$ (the merger remnant's disc's angular momentum at $t_\text{post/eq}$), and $\vec{j}_\text{orb}$ (the interaction's orbital angular momentum, calculated by $\vec{r}_\text{fp} \times \vec{v}_\text{fp}$, with $\vec{r}_\text{fp}$ and $\vec{v}_\text{fp}$ being the relative position and velocity of the secondary galaxy's centre with respect to the primary galaxy's centre at the first periapsis $t_\text{fp}$). To make sure that $\vec{j}_\text{post/eq}$ represents the disc orientation, we only summed up the angular momentum of the stars assigned to the disc component (as in Fig.~\ref{fig:disk_decomposition}) according to our kinematic decomposition (Section~\ref{subsect:disk_formation}). According to Fig.~\ref{fig:angular_momentum_direction_change}, $\vec{j}_\text{post/eq}$ aligns very well with $\vec{j}_\text{orb}$, with the angle between them ($\Delta\theta_\text{e-o}$) being less than~$30^{\circ}$ for all nine codes. On the other hand, the rotational axis of the main galaxy at $t_\text{start}$ ($\vec{j}_\text{start}$) is distinct from the rotational axis of the merger remnant's disc ($\vec{j}_\text{post/eq}$), as there is a large angle ($\Delta\theta_\text{e-s} \geq 38^{\circ}$, with the majority exceeding $ 80^{\circ}$) between the two. 
Noticeably, even for GADGET-3 and GADGET-4, whose main galaxy has a disc before the \textit{target} merger, the new disc still follows the orientation of the merger's orbit rather than the orientation of the old disc. The orientation between the pre-merger disc and the post-merger disc in these two cases is larger than $90^{\circ}$, showing that the \textit{target} merger considerably changes the orientation of the pre-merger disc. This change in the disc orientation in GADGET-3 and GADGET-4 can also be seen in Fig.~\ref{fig:gas_projection_of_mergers}, where the disc at $t_\text{start}$ is oriented differently from the disc at $t_\text{cls}$ (slightly edge-on at $t_\text{start}$ but directly face-on at $t_\text{cls}$) even though we aligned the two projection plots using the same normal axis. Indeed, the \textit{target} merger may even destroy the pre-merger disc and turn it into a part of the post-merger bulge through violent relaxation \citep{Lynden-Bell+1967, Hopkins+2009}, as the circularity distribution of the old star population is largely centred around zero for both GADGET-3 and GADGET-4. Thereby, for the \textit{target} merger, the disc of the merger remnant is created from the interaction's orbit and is kinematically separate from the pre-merger disc.

In summary, regardless of the code architectures and feedback models, we find that the orbital orientation of the \textit{target} merger helps shape the orientation of the merger remnant's disc. The secondary galaxy brings the gas in with an orbital angular momentum, and this gas accretes onto the primary galaxy and forms stars while still preserving that angular momentum direction. When examining Milky Way-mass galaxies from the TNG-50 simulation, \cite{Bell+2026} also arrived at a similar conclusion that most galaxies show alignment with the orbital angular momentum of their most massive merger progenitor. Hence, the rotational axis of a galaxy's post-merger disc can act as a kinematic record of the trajectory of the galaxy's past merging events.  

\section{Discussion}
\label{sec:discussion}

\subsection{Comparison with Santa Cruz semi-analytic models}
\label{sec:comparison_sam}

\begin{figure*}[tbh]
    \centering
    \includegraphics[width=\linewidth]{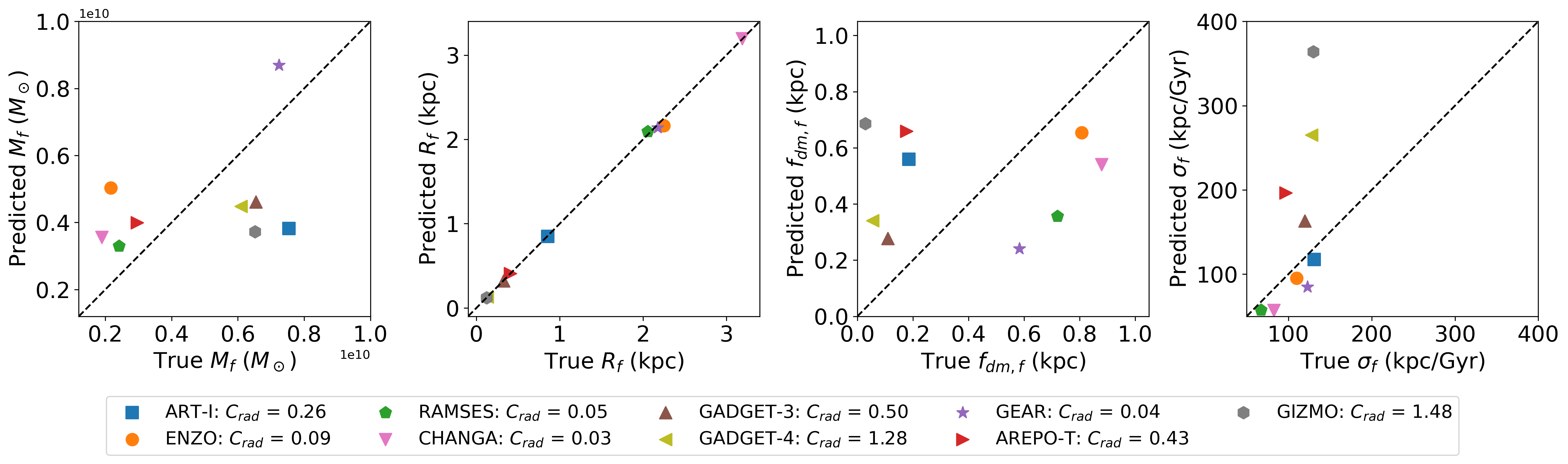}
    \caption{Comparison of the merger remnant's properties (evaluated at the coalescence timestep) between being directly obtained from the \texttt{CosmoRun} simulations and being predicted using \citet{Covington+2008}'s semi-analytic model. From left to right, the compared properties are stellar mass ($M_{f}$), stellar 3D half-mass radius ($R_{f}$), DM fraction inside $0.5R_{f}$ ($f_\text{dm,f}$), and stellar velocity dispersion inside $R_{f}$ ($\sigma_{f}$). For each code, the value of $C_\text{rad}$, the parameter to adjust the weighting of the radiated energy loss in the SAM, is selected to minimise the error in $R_{f}$. The model overall agrees well with the simulations regarding the stellar mass and stellar half-mass radius, while struggling to predict the DM fraction and velocity dispersion for codes with a very compact merger remnant.}
    \label{fig:compare_with_SAM_models}
\end{figure*}

In addition to direct hydrodynamic simulations, attempts to study galaxy mergers were also made using semi-analytic models (SAMs), where complex baryonic processes are parameterised and modelled on top of N-body DM simulations. One such attempt is the Santa Cruz SAM with the framework to predict galaxy merger outcomes developed by \cite{Covington+2008} and later extended by \cite{Porter+2014}. Taking the progenitor galaxies' properties (total mass, stellar mass, half-mass radius, internal kinetic energy, gas fraction) and orbital parameters (initial distance and velocity) as inputs, \cite{Covington+2008} constructed a physical, dissipative model that is calibrated with idealised hydrodynamic simulations to predict the stellar mass ($M_{f}$), stellar half-mass radius ($R_{f}$), DM fraction within $0.5R_{f}$ ($f_\text{dm, f}$), and stellar velocity dispersion within $R_{f}$ ($\sigma_{f}$) of the merger remnant.  In this section, we compare our \texttt{CosmoRun} results of the \textit{target} mergers with the Santa Cruz SAM to assess whether the SAM's predictions agree with simulations across different hydrodynamic solvers and stellar feedback prescriptions.

Fig.~\ref{fig:compare_with_SAM_models} shows the comparison between the properties of the merger remnants obtained from the \texttt{CosmoRun} simulations (evaluated at the coalescence timestep $t_\text{cls}$) and predicted from the Santa Cruz SAM. In addition to the initial properties of the merger, the model also employs a set of tunable parameters for calibration (see Table B2 of \citealp{Covington+2008}). An important parameter among them is $C_\text{rad}$, which sets the relative weighting of radiated energy that is lost from the merging interaction. Among the four variables predicted by the SAM, tuning $C_\text{rad}$ directly affects the predicted stellar half-mass radius $R_{f}$ and influences the predicted velocity dispersion $\sigma_{f}$ as a secondary effect. Adjusting $C_\text{rad}$ does not change the predicted stellar mass $M_{f}$ or the central DM fraction $f_\text{dm, f}$. This parameter can be adjusted depending on the type of mergers, in which more dissipationless (dissipative) mergers have a lower (higher) $C_\text{rad}$ value \citep{Porter+2014}. Because each of the \texttt{CosmoRun} simulations uses a different stellar feedback scheme that can influence the radiative energy loss, for each code, we chose the $C_\text{rad}$ that minimises the error between the model's prediction and the simulation's outcome on $R_{f}$ (i.e., minimizing $(R_\text{f,predicted} - R_\text{f,sim})^2/R_\text{f,sim}^2$). The adopted $C_\text{rad}$ values are listed in the legend of Fig.~\ref{fig:compare_with_SAM_models}. Our computed values of $C_\text{rad}$ range from 0.03 to 1.48, which are still well inside the range of the calibrated values found in \cite{Porter+2014} when they tested the SAM with binary merger simulations. For other tunable parameters in the model, we followed the default values reported in \cite{Covington+2008}.

According to Fig.~\ref{fig:compare_with_SAM_models}, the merger remnant's stellar mass obtained from the semi-analytic model agrees well with the stellar mass from our simulations without the need to change any adjustable parameters from the model. The agreement is within~0.5 dex, with most of the codes reaching a difference of only~0.3 dex between the predicted and the true stellar mass. Regarding the stellar half-mass radius $R_{f}$, by tuning the parameter $C_\text{rad}$, the model's predicted values can match the true values for all codes. Even though this agreement is achieved by fine-tuning, the tuned values of $C_\text{rad}$ can help imply the level of dissipation or energy loss that each simulated merger has. In \texttt{CosmoRun}, codes with kinetic feedback (ART-I, GADGET-3, GADGET-4, and GIZMO) have a higher $C_\text{rad}$ compared to codes that do not use kinetic feedback. In the SAM, $C_\text{rad}$ controls the relative weight of the radiated energy. Given that the stellar mass is well predicted by the model, we can argue that the codes with kinetic feedback are more effective in dissipating energy during the \textit{target} merger. Because gas dissipates energy more effectively, it can cool and collapse into the galactic center more easily, which helps explain why these codes experience an earlier starburst and form a more compact remnant (see Paper~IX - Part~1 for a more in-depth discussion). For CHANGA, RAMSES, and GEAR, similar to our arguments in Sections~\ref{subsect:radial_distribution} and ~\ref{subsect:burstfraction_vs_compaction}, their delayed cooling scheme and the strong preventative feedback from the superbubble scheme prevent radiative losses and gas cooling in the galaxies. This inefficient gas cooling explains why these three codes have the lowest $C_\text{rad}$ values. ENZO also has a relatively low $C_\text{rad}$ compared to AREPO-T, which also uses only thermal feedback for its stellar feedback prescription. The reason can be that ENZO has more energy output per supernova event compared to AREPO-T ($5\times10^{52}$ versus $5\times10^{51}$ ergs), which causes more gas heating and prevents gas cooling more efficiently. A future AGORA paper (Lindh et al., in prep.) will explore the effect of different feedback schemes on the CGM, which can shed more light on this discrepancy between ENZO and AREPO-T, even though they use the same feedback group.

Even though the model predicts $M_{f}$ and $R_{f}$, it is limited in its predictions for $f_\text{dm, f}$ and $\sigma_{f}$. In particular, we find that the model overestimates $f_\text{dm, f}$ for codes that experience strong compaction after the merger, while underestimating $f_\text{dm, f}$ for codes whose remnants expand. This discrepancy is due to two reasons. First, the estimation of $f_\text{dm, f}$ depends on the tunable parameter $C_\text{stars}$, which is the fraction of stellar mass over total mass inside one-half of $R_f$. $C_\text{stars}$ is chosen to be 0.35 in \cite{Covington+2008}'s fiducial model. We attempted to tune this parameter to replicate the simulated $f_\text{dm, f}$, and find that for ENZO, RAMSES, CHANGA, and GEAR, a value around $0.06\text{--}0.15$ will help fit $f_\text{dm, f}$ better. On the other hand, for ART-I, GADGET-3, GADGET-4, AREPO-T, and GIZMO, we find that $C_\text{stars}$ needs to be larger than 1 for the SAM to match the simulations, which is unphysical because $C_\text{stars}$ is a ratio. Therefore, adjusting $C_\text{stars}$ by itself does not help address the discrepancy. The second reason for the SAM's inaccurate estimate of $f_\text{dm, f}$ is the model's assumption that the inner region (within half of the stellar half-mass radius) of the merger remnant contains the same amount of DM mass as the sum of the inner regions of the two progenitors \citep{Covington+2008}. For brevity, we refer to the DM mass within half of the stellar half-mass radius ($0.5R_f$) as just "inner DM mass". We evaluated this assumption using our simulations and find that it becomes less accurate for codes whose remnants change substantially in size. In our analysis, the assumption breaks down the most for codes exhibiting strong compaction (ART-I, AREPO-T, and GIZMO). After a compaction event, the stellar half-mass radius of the galaxy decreases considerably, making the inner DM mass of the remnant $1\text{--}2$ dex smaller than the sum of the progenitors' initial inner DM masses. Even for codes whose stellar half-mass radius changes only slightly after the merger (ENZO, GADGET-3, and GADGET-4), the remnant's inner DM mass is still reduced. On the other hand, for codes whose main galaxy expands (RAMSES, CHANGA, and GEAR), the larger stellar half-mass radius of the remnant yields an inner DM mass higher by a factor of $1.3\text{--}2$ compared to the total initial inner DM mass. In summary, the SAM tends to overestimate $f_{\rm dm,f}$ for codes undergoing strong compaction (whose assumed inner DM mass of the remnant is too high) and to underestimate $f_{\rm dm,f}$ for codes undergoing strong expansion (whose assumed inner DM mass of the remnant is too low).

Given that the predicted $R_f$ is tuned to match the simulated value and the predicted $M_f$ is consistent with the simulations, when the model over-predicts (under-predicts) $f_\text{dm, f}$, it consequently overestimates (underestimates) the velocity dispersion because $\sigma_f$ and $f_\text{dm, f}$ are related by $\sigma_f^2 = C_{\mathrm{vir}} \, \frac{G M_f}{R_f\left(1 - f_{\mathrm{dm},f}\right)}$ \citep{Covington+2008}. Therefore, even though the Santa Cruz SAM performs well in predicting the stellar mass and stellar half-mass radius of our merger remnants, its calculations of the central DM fraction and velocity dispersion need more calibration to reproduce the results of cosmological hydrodynamic simulations, especially when a merger remnant undergoes substantial compaction or expansion in size.

\subsection{Comparison with observations}

Our finding that a major merger can drive compaction and disc formation is broadly consistent with the observational results. As post-starburst galaxies (PSBs) are often interpreted as galaxies in a transitional phase following a merger-driven starburst \citep{Sazonova+2021, Verrico+2023}, they serve as proxies for merger remnants. Observations of PSB galaxies reveal that they typically display smaller effective radii and are more compact than coeval star-forming and quiescent galaxies at approximately similar stellar mass \citep{Whitaker+2012, Almaini+2017, Setton+2022}. Observational studies also point out that galaxies undergoing a merger have a stronger central starburst and excess of central stellar mass compared to isolated galaxies with similar global SFR enhancements \citep{Thorp+2024, Ellison+2026}. These results regarding compaction and central star formation are consistent with our findings for ART-I, ENZO, GADGET-3, GADGET-4, AREPO-T, and GIZMO, where we see that stars formed during the coalescence and post-coalescence stage of the \textit{target} merger are much more concentrated in the inner radius (Figs.~\ref{fig:radial_star_distribution} and \ref{fig:circularity_vs_distanc}).

Our result that a stellar disc can form after a major merger event is also supported by several observational studies, where a rotational structure exists in a merger remnant. \cite{Hammer+2009} examined J033245.11-274724.0, a luminous IR galaxy at $z = 0.43$ with a warped, fast-rotating disc and two spiral arms, and concluded that the galaxy properties are best explained as the remnant of a major merger of two gas-rich, equal-mass galaxies. From CO imaging of nearby merger remnants, \cite{Ueda+2014} reported that~80\% (24/30) of their sample show evidence of rotating molecular gas discs. In a study of superluminous spiral galaxies drawn from SDSS, \cite{Ogle+2016} found four systems in the late stage of major mergers and proposed that such superluminous spirals may originate from gas-rich, spiral–spiral major mergers. In contrast, statistical studies using PSBs show that PSBs are more bulge-dominated with large Sérsic indices \citep{Quintero+2004, Mendel+2013, French+2021a}.

There are some caveats regarding the comparison between our results and observations. First, the feedback dependence we recover has no direct observational counterpart, since observations cannot isolate a single merger under a controlled feedback prescription. What the observational data reveal instead is a statistical result, which means that there is also a population of merger remnants that are not compact or forming a disc. In other words, all nine remnants of our \textit{target} merger in \texttt{CosmoRun} can be regarded as valid simulation realizations. Furthermore, we caution that our comparison using PSBs is relative, as most of the PSBs sample is at $z < 2$, while our \textit{target} merger is at $z \approx 4.5$. Also, unlike the PSBs case, none of our simulated galaxies are quenched after the \textit{target} merger, as the high gas fraction and accretion at high redshift help sustain the fuel for star formation even after strong starburst events. Lastly, the mass regimes in the observational studies and in this work are also significantly different because of the difference in redshift range. The stellar mass range of \texttt{CosmoRun} galaxies at the time of the \textit{target} merger is about $10^{9}\text{--}10^{10} \text{M}_\odot$, while the galaxies in the aforementioned observational work are about $10^{10}\text{--}10^{11.5} \text{M}_\odot$.

\section{Conclusion}
\label{sec:conclusion}

Building on the analysis of Paper~IX - Part~1, this paper extends our investigation of a major galaxy merger at $z \approx 4.5$ in the AGORA \texttt{CosmoRun} cosmological hydrodynamic simulation suite. We compare the merger's morphological impact on the main galaxy across nine state-of-the-art simulation codes. The power of the AGORA comparison project is the robust calibration between the codes. This calibration allows the differences in the merger remnant's morphology to be attributed to the stellar feedback schemes and code architectures rather than to numerical artifacts. Our main results are as follows:

\begin{itemize}

    \item The codes' adopted stellar feedback schemes affect the main galaxy's morphological transformation after the merger. Codes with kinetic feedback experience compaction and disc formation earlier in the interaction (during the infall and the first passage stages), while codes using only thermal feedback experience them later in the coalescence and post-coalescence stages. In contrast, codes that employ the delayed cooling or the superbubble scheme with only thermal feedback do not undergo either compaction or disc formation during the major merger's timescale. (Figs.~\ref{fig:radial_star_distribution}, \ref{fig:circularity_distribution},~\ref{fig:disk_decomposition}, and \ref{fig:circularity_vs_distanc}) 

    \item The change in the stellar half-mass radius is negatively correlated with the merger's burst fraction. The merger remnant becomes more compact and decreases in the stellar half-mass radius in codes that experience a higher burst fraction ($f_\text{sb} > 0.7$). On the other hand, the merger remnant grows in size in codes with a low burst fraction ($f_\text{sb} < 0.6$). Codes using the delayed cooling feedback with long delay time (RAMSES and GEAR) do not follow this correlation. An important caveat is that this correlation is measured across a limited number of different simulation codes, and a large sample of mergers within a single code is necessary to robustly establish the correlation. (Fig.~\ref{fig:HalfMassRadiusRatio_vs_burstFraction})

    \item For all codes, we find that the rotational angular momentum axis of the merger remnant's disc aligns with the orbital angular momentum axis of the merging interaction rather than the pre-merger rotational angular momentum axis. This result suggests that the infalling gas, carried by the secondary galaxy, retains its orbital angular momentum as it accretes onto the primary galaxy to form a disc of the merger remnant. (Fig.~\ref{fig:angular_momentum_direction_change})

    \item The stellar mass and stellar half-mass radius of the merger remnants in all \texttt{CosmoRun} codes are in reasonable agreement with the predictions of the Santa Cruz semi-analytic model of galaxy mergers. Nonetheless, the semi-analytic model has difficulty predicting the central DM fraction and velocity dispersion of the merger remnants, particularly for codes whose main galaxies undergo considerable compaction or expansion. (Fig.~\ref{fig:compare_with_SAM_models})
    
\end{itemize}

Even though this study has a limitation of relying on only one merger in each code for the comparison, its findings nonetheless underscore the sensitivity of galaxy mergers to stellar feedback models in cosmological hydrodynamic simulations. The adopted feedback scheme of a cosmological simulation may bias the predicted morphological transformation of galaxy mergers. This feedback-induced bias in the morphological response of individual galaxy mergers can propagate into the predicted demographics of simulated galaxy morphology and, in turn, into our understanding of galaxy evolution from simulations.

\section*{ACKNOWLEDGEMENTS}

We thank all of our colleagues in the AGORA Collaboration for their collaborative and supportive spirit, which has allowed the collaboration to remain strong as a platform to foster and launch multiple science-oriented comparison efforts. THN acknowledges support from the National Center for Supercomputing Applications Center and its Center for Astrophysical Surveys. THN and KSSB acknowledge the University of Illinois at Urbana-Champaign for their continued support, and the support of the Delta Supercomputer as well as the ACCESS program for computing grants PHYS240175 and PHY250173. We thank Jinning Liang for sharing the code of their work on kinematic decomposition publicly on GitHub.
R.R.C.acknowledges financial support from the Spanish Ministry of Science and Innovation through the research grants: PID2021-123417OBI00, funded by MCIN/AEI/10.13039/501100011033/FEDER, EU; PCI2022-135023-2, funded by MCIN/AEI/10.13039/ 501100011033; the EU “NextGenerationEU” / PRTR; and PID2024-157374OBI00, funded by MICIU/AEI/10.13039/501100011033/FEDER, EU; and the IND2022/TIC-23643 project funded by Comunidad de Madrid. 
J.-H.K.’s work was supported by the National Research Foundation of Korea (NRF) grant funded by the Korea government (MSIT; No. 2022M3K3A1093827 and No. 2023R1A2C1003244). His work was also supported by the National Institute of Supercomputing and Network/ Korea Institute of Science and Technology Information with supercomputing resources including technical support, grants KSC-2022-CRE-0355 and KSC-2024-CRE-0232. His work was also supported by the GlobalLAMP Program of the NRF grant funded by the Ministry of Education (No. RS-2023-00301976). 
KN acknowledges support from JSPS KAKENHI grant 20H00180, 24H00002, 24H00241, JP25K01032, and the JSPS International Leading Research (ILR) project, JP22K21349.
KN also acknowledges support from the Kavli IPMU, the World Premier Research Center Initiative (WPI), UTIAS, the University of Tokyo.  
DC is supported by research grant  PID2024-156100NB-C21 financed by MICIU/AEI /10.13039/501100011033 / FEDER, EU., and the research grant
CNS2024-154550 funded by MI-CIU/AEI/10.13039/501100011033.  
HV was supported by a grant UNAM-PAPIIT-IN111425.

\section*{DATA AVAILABILITY}
The AGORA \texttt{CosmoRun} raw simulation snapshots from Papers~\citetalias{Roca-Fabrega+2021} and \citetalias{Roca-Fabrega+2024} are publicly available at \url{https://flathub.flatironinstitute.org/agora}. The analysis codes and the simulation metadata underlying this project can be shared on reasonable request to the corresponding authors and the AGORA collaboration.

%

\vspace{5mm}





\appendix

\section{Disc Decomposition Method for GEAR}
\label{subsect:GEAR_Disk_Decomposition}

\begin{figure*}[tbh]
    \centering
    \includegraphics[width=0.9\linewidth]{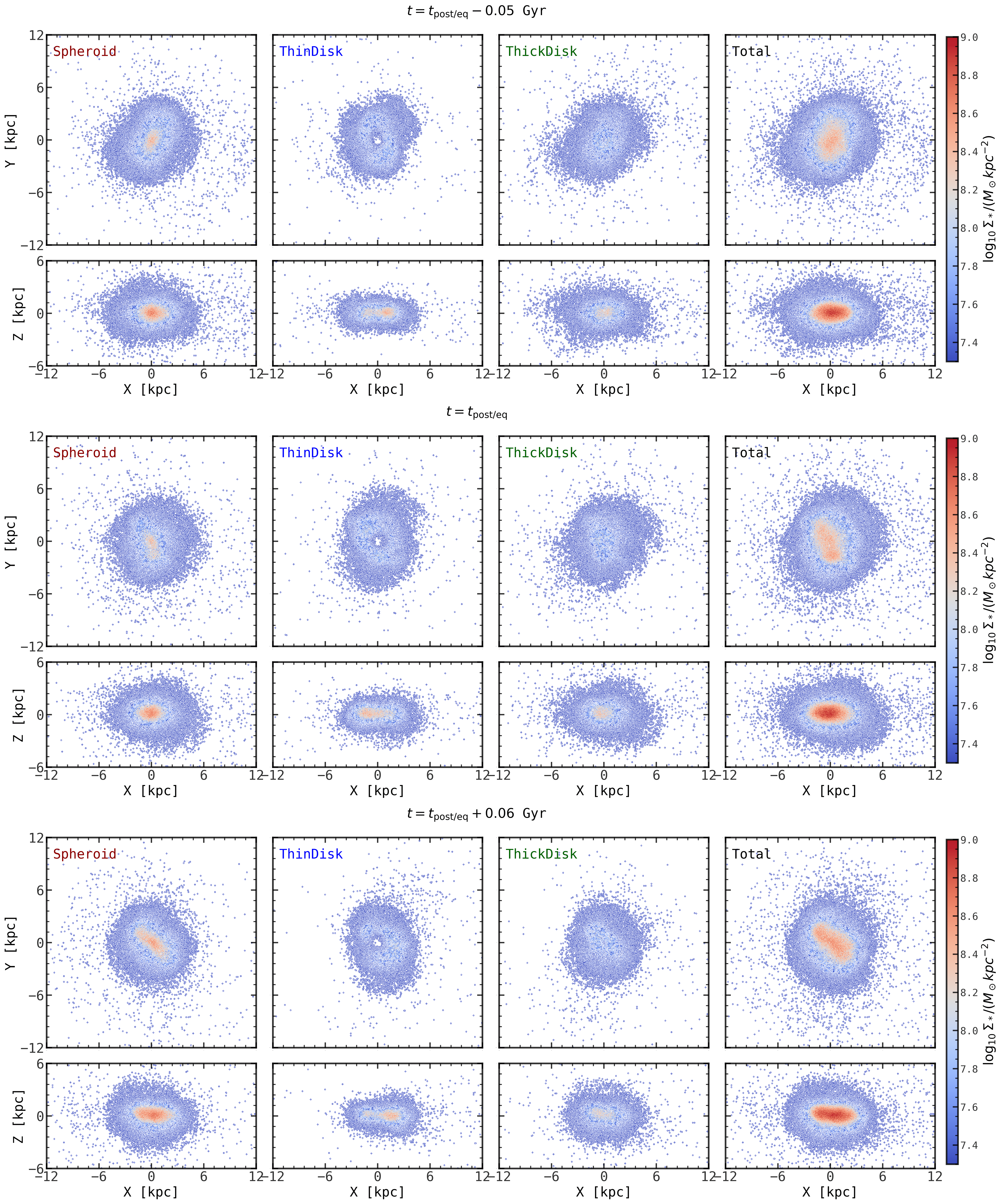}
    \caption{Face-on and edge-on maps of stellar surface density for different kinematic components at three timesteps around $t_\text{post/eq}$. The kinematic decomposition follows the method of \cite{Liang+2025}. From left to right, each set of columns represents the spheroid, thin disc, thick disc, and total stellar component. Since the merging galaxies have not coalesced, we notice two stellar nuclei orbiting around each other at a very close distance (< 2 kpc). This behaviour is clearest in the edge-on thin-disc views, in which one of the stellar nuclei oscillates from left (panel $t_\text{post/eq} - 0.5$ Gyr) to right (panel $t_\text{post/eq}$) and back to left (panel $t_\text{post/eq} + 0.6$ Gyr). This orbital motion contaminates the bulge-disc decomposition. The disc component instead constitutes the two orbiting nuclei, while the mean circularity of the spheroid is shifted above zero (as seen in Fig.~\ref{fig:GEAR_alternative_disk_decomposition_method}).}
    \label{fig:decomposition_GEAR_fluctuation}
\end{figure*}

\begin{figure*}[tbh]
    \centering
    \includegraphics[width=\linewidth]{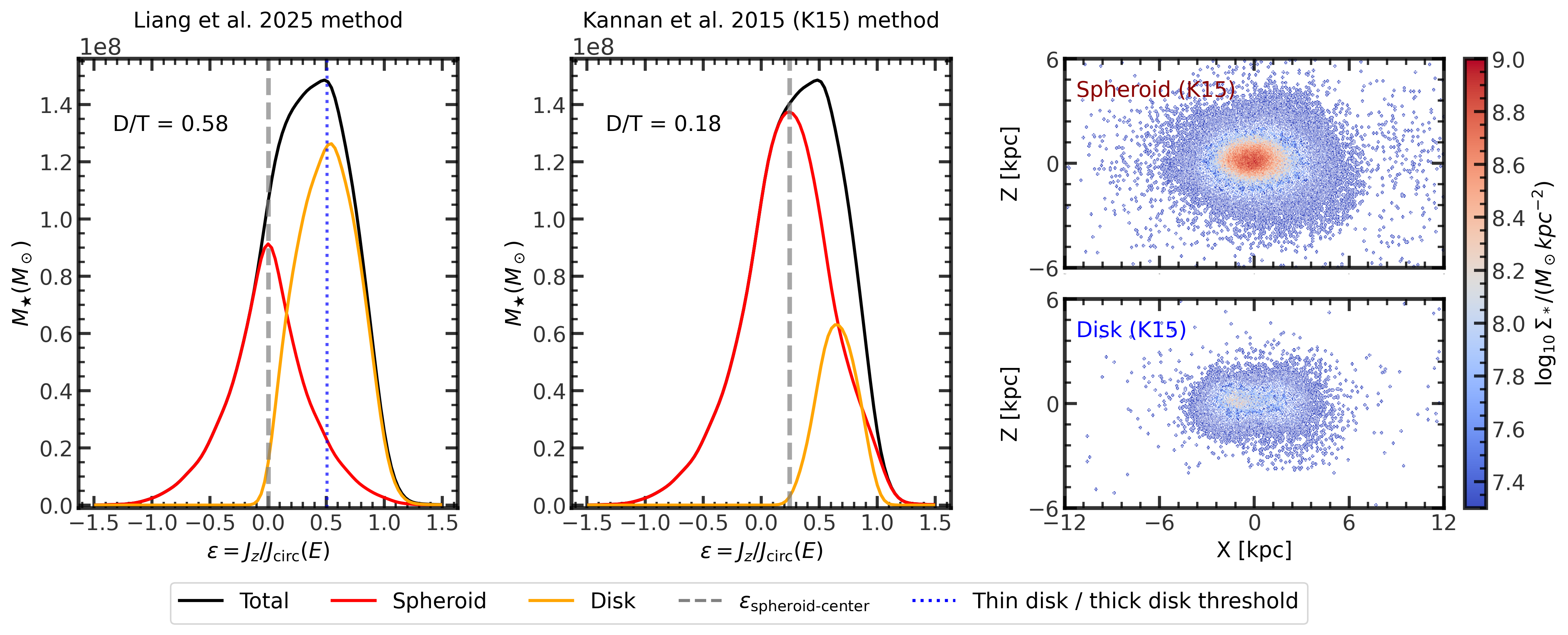}
    \caption{Left and Middle: the morphological decomposition between the \citet{Liang+2025}'s method and the \citet{Kannan+2015}'s method (K15). Right: the edge-on maps of stellar surface density for the spheroid and the disc components decomposed by \cite{Kannan+2015}'s method. Because the galaxy system in GEAR undergoes a significantly prolonged coalescence stage (compared to other codes) during the \textit{target} merger, at $t_\text{post/eq}$, the stellar circularity distribution is biased towards orbital motion between the two nuclei instead of reflecting the disc's rotational motion, leading to an artificially high D/T value. Therefore, we decided to adopt the K15 kinematic decomposition method, which does not assume that the spheroid's $\epsilon$ distribution is symmetric around zero. The updated disc component is no longer contaminated by the two orbiting stellar nuclei as in Fig.~\ref{fig:decomposition_GEAR_fluctuation}.}
    \label{fig:GEAR_alternative_disk_decomposition_method}
\end{figure*}

In Section~\ref{subsect:disk_formation}, we employed the method by \citet{Liang+2025} to decompose the stellar circularity distribution into the spheroid and the disc component. In all codes but GEAR, this method performs well as the two merging galaxies fully coalesce at $t_\text{post/eq}$ into a single-nucleus system. On the other hand, the two stellar cores (or stellar nuclei) in GEAR orbit each other for a considerably long time with a relatively small distance (< 2kpc, fig.12 of Paper~IX - Part~1). This orbital motion interferes with the adopted kinematic decomposition method. When two nuclei that are comparable in size orbit each other, the centre of mass of the system will be between the two nuclei, and the z-axis angular momentum of the system will be dominated by the orbital motion instead of the internal rotation of each galaxy. In addition, as the distance between the two nuclei is small, it is challenging to completely remove one galaxy's motion without removing the intrinsic disc/bulge of another galaxy. As a result, the stellar circularity distribution of the system is not symmetric about zero but instead centred on $\epsilon \approx 0.2$ because of the nuclei pair's orbital angular momentum, as shown in Fig.~\ref{fig:circularity_distribution}. This offset compromises the application of the method by \citet{Liang+2025}, which assumes that the spheroidal component is centred at $\epsilon = 0$ and the remaining component is attributed to the disc. 

Figs.~\ref{fig:decomposition_GEAR_fluctuation} and ~\ref{fig:GEAR_alternative_disk_decomposition_method} (left subplot) display our attempt to apply \citet{Liang+2025}'s decomposition technique on GEAR's galaxy system at $t_\text{post/eq}$. After the kinematic decomposition, two stellar nuclei are visible in orbit at a small separation ($< 2$~kpc) in the edge-on maps of the thin-disc component, showing that the disc component is contaminated. Within $\approx 0.1$ Gyr around the equivalent timestep, we see that one nucleus oscillates from the right side of the plot (second column, top panel, at $t_\text{post/eq} - 0.5$ Gyr) to the left (second column, middle panel, at $t_\text{post/eq}$) and then back to the right side again (second column, bottom panel, at $t_\text{post/eq} + 0.6$ Gyr), demonstrating a two-body obit. Both stellar nuclei are likewise apparent in the face-on view of the total stellar surface-density map (rightmost column).

Because the total stellar angular momentum used to define the disc plane is dominated by the stellar nuclei's orbital motion rather than the galaxy's intrinsic rotation, the orbital motion imprints rotation on otherwise pressure-supported motion, hence displacing the peak of the spheroid distribution ($\epsilon_\text{spheroid-centre}$) away from zero. Therefore, in this particular case, it is not justified to assume that the spheroid's $\epsilon$ distribution is symmetric about zero. As shown in the left subplot of Fig~\ref{fig:GEAR_alternative_disk_decomposition_method}, \citet{Liang+2025}'s assumption underestimates the spheroid fraction and hence exaggerates the D/T value, leading to an unrealistic, non-uniform disc component that is contaminated by the two nuclei's orbit. Therefore, we applied a different morphological decomposition method by \cite{Kannan+2015}. In this method, $\epsilon_\text{spheroid-center}$ is determined by using (1) the local maxima of the $\epsilon$ distribution function or (2) the inflection point where the second derivative turns from negative to positive, if the bulge and the disc distributions merge and the local maxima vanish. This search spans only $-0.3 \leq \epsilon\leq 0.3$. If the peak is not found, the galaxy is considered bulgeless. The decomposition result is shown in the middle and the right subplots of Fig~\ref{fig:GEAR_alternative_disk_decomposition_method}, with $\epsilon_\text{spheroid-centre}$ computed to be $\approx 0.24$. The new $\epsilon_\text{spheroid-centre}$ value helps capture the spheroid component more accurately. As shown in the edge-on stellar surface density maps, with \cite{Kannan+2015}'s method, the disc component is now more uniform and is not contaminated by the orbital motion of the two stellar nuclei anymore. In fact, the two nuclei are assigned to be inside the spheroid component of the galaxy system. As a result, the D/T value becomes 0.18, which means that the system is not a disc galaxy. This D/T value is more consistent with the visual inspection of the galaxy (Fig.~\ref{fig:gas_projection_of_mergers}). It is also important to note that this issue does not happen with other codes, and it is still appropriate to use \citet{Liang+2025}'s method in other cases. We do not detect any orbiting stellar nuclei in other codes at $t_\text{post/eq}$, and the local maxima or the inflection point of the $\epsilon$ distribution are located around zero, as shown in Fig.~\ref{fig:circularity_distribution}.


\bibliography{sample631}{}
\bibliographystyle{aasjournal}



\end{document}